\documentclass[reprint,two-column,double-spaced,groupedaddress,superscriptaddress,pra]{revtex4-2}

\usepackage[normalem]{ulem}

\usepackage{mathrsfs}
\usepackage{enumerate}%numbering
\usepackage{graphicx}% Includeure files
\usepackage{physics}%ketbra

\usepackage[colorlinks]{hyperref}%hyperlink
\usepackage{tensor}
\usepackage{leftidx}
\usepackage{braket}
\usepackage[dvipsnames]{xcolor}

\makeatletter
\def\@fnsymbol#1{\ensuremath{\ifcase#1\or \dagger\or*\or \ddagger\or
   \mathsection\or \mathparagraph\or \|\or **\or \dagger\dagger
   \or \ddagger\ddagger \else\@ctrerr\fi}}
    \makeatother

\begin{document}

\title{Quantum correlation generation capability of experimental processes}

\author{Wei-Hao Huang}
\thanks{These authors contributed equally to this work.}
\affiliation{Department of Engineering Science, National Cheng Kung University, Tainan 70101, Taiwan}
\affiliation{Center for Quantum Frontiers of Research $\&$ Technology, National Cheng Kung University, Tainan 70101, Taiwan}

\author{Shih-Hsuan Chen}
\thanks{These authors contributed equally to this work.}
\affiliation{Department of Engineering Science, National Cheng Kung University, Tainan 70101, Taiwan}
\affiliation{Center for Quantum Frontiers of Research $\&$ Technology, National Cheng Kung University, Tainan 70101, Taiwan}

\author{Chun-Hao Chang}
\thanks{These authors contributed equally to this work.}
\affiliation{Department of Engineering Science, National Cheng Kung University, Tainan 70101, Taiwan}
\affiliation{Center for Quantum Frontiers of Research $\&$ Technology, National Cheng Kung University, Tainan 70101, Taiwan}

\author{Tzu-Liang Hsu}
\affiliation{Department of Engineering Science, National Cheng Kung University, Tainan 70101, Taiwan}
\affiliation{Center for Quantum Frontiers of Research $\&$ Technology, National Cheng Kung University, Tainan 70101, Taiwan}

\author{Kuan-Jou Wang}
\affiliation{Department of Engineering Science, National Cheng Kung University, Tainan 70101, Taiwan}
\affiliation{Center for Quantum Frontiers of Research $\&$ Technology, National Cheng Kung University, Tainan 70101, Taiwan}

\author{Che-Ming Li}
\email{cmli@mail.ncku.edu.tw}
\affiliation{Department of Engineering Science, National Cheng Kung University, Tainan 70101, Taiwan}
\affiliation{Center for Quantum Frontiers of Research $\&$ Technology, National Cheng Kung University, Tainan 70101, Taiwan}

\date{\today}

\begin{abstract}
Einstein-Podolsky-Rosen (EPR) steering and Bell nonlocality illustrate two different kinds of correlations predicted by quantum mechanics. They not only motivate the exploration of the foundation of quantum mechanics, but also serve as important resources for quantum-information processing in the presence of untrusted measurement apparatuses. Herein, we introduce a method for characterizing the creation of EPR steering and Bell nonlocality for dynamical processes in experiments. We show that the capability of an experimental process to create quantum correlations can be quantified and identified simply by preparing separable states as test inputs of the process and then performing local measurements on single qubits of the corresponding outputs. This finding enables the construction of objective benchmarks for the two-qubit controlled operations used to perform universal quantum computation. We demonstrate this utility by examining the experimental capability of creating quantum correlations with the controlled-phase operations on the IBM Quantum Experience and Amazon Braket Rigetti superconducting quantum computers. The results show that our method provides a useful diagnostic tool for evaluating the primitive operations of nonclassical correlation creation in noisy intermediate scale quantum devices.
\end{abstract}

\maketitle

\section{\label{sec:introduction}Introduction}

Quantum computation relies on the properties of quantum mechanics to achieve computational speeds that are unattainable with ordinary digital computing~\cite{feynman2018simulating, deutsch1985quantum, lloyd1996universal, divincenzo2000physical, nielsen2002quantum, georgescu2014quantum}. Quantum technologies have now become so well developed that they have enabled the building of intermediate-size computers with 50-100 qubits. However, the noise produced in the quantum gates of such computers limits the quantum results obtained, and the error rate may be so high as to be almost impossible to evaluate. As a result, these computers have come to be known as noisy intermediate-scale quantum (NISQ) devices~\cite{preskill2018quantum, bruzewicz2019trapped, kjaergaard2020superconducting}.
Various physical platforms have been presented for the implementation of quantum computers, ranging from silicon-based systems~\cite{zwanenburg2013silicon, kane1998silicon, ladd2002all} to superconducting circuits~\cite{madsen2022quantum, xiang2013hybrid} constructed with Josephson junctions~\cite{makhlin2001quantum, nakamura1999coherent, krantz2019quantum} and trapped ions using laser pulses~\cite{haffner2008quantum, cirac1995quantum}.
Thanks to these developments, various quantum computers have now been commercialized, including the IBM Quantum (IBM Q) Experience~\cite{ibmq} and Amazon Web Services (AWS) Amazon Braket~\cite{aws}.

However, it is essential to go beyond simply boosting the number of physical qubits on a quantum processor to enable quantum information processing (QIP)~\cite{knill2008randomized}. For example, benchmarking different physical platforms typically considers the entire range of critical parameters that affect qubits and influence the current and future capabilities of the quantum processor architecture~\cite{cross2019validating, jurcevic2021demonstration, blume2013robust, blume2017demonstration, benedetti2019generative, chen2014qubit, huang2019fidelity, xue2019benchmarking, veldhorst2014addressable}. Typically, these parameters include not only performance metrics, such as the qubit connectivity, overall gate speed, and gate fidelity, but also the qubit manufacturability. In the short term, it may be sufficient to develop practical benchmarks across different platforms based simply on the figures of merit demonstrated by certain applications~\cite{knill2008randomized, cross2019validating, knill2001benchmarking, wright2019benchmarking, mccaskey2019quantum, harrigan2021quantum}. However, to avoid optimizing or tuning quantum devices such that their performance is maximized only under certain specific benchmarking methods, it is desirable to develop a standardized set of algorithmic benchmarks based on a wider consideration of multiple resource constraints~\cite{cross2019validating}.

Quantum correlations, such as quantum entanglement~\cite{einstein1935can, horodecki2009quantum}, Einstein-Podolsky-Rosen (EPR) steering~\cite{schrodinger1935discussion, wiseman2007steering, uola2019quantum}, and Bell nonlocality~\cite{bell1964einstein, clauser1969proposed, brunner2014bell}, can be considered as such resource constraints for developing benchmarks. These resources have been successfully demonstrated for various applications, including quantum teleportation~\cite{bennett1993teleporting, bouwmeester1997experimental}, quantum secret sharing~\cite{hillery1999quantum, chen2005experimental}, and one-way quantum computing~\cite{PhysRevLett.86.5188, raussendorf2003measurement, walther2005experimental} (entanglement); one-sided device-independent quantum key distribution~\cite{branciard2012one} and randomness certification~\cite{uola2019quantum} (EPR steering); and quantum cryptography~\cite{ekert1991quantum}, communication complexity problem~\cite{brunner2014bell, brukner2004bell, buhrman2010nonlocality}, and device-independent QIP~\cite{brunner2014bell, de2014nonlocality} (Bell nonlocality).

\begin{figure*}[t]
\includegraphics[width=17.8cm]{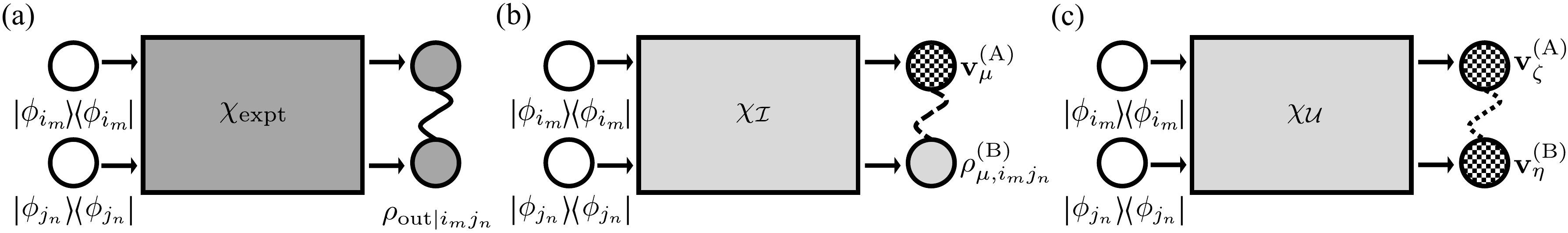}
\caption{Quantifying the capability of an unknown process for creating quantum correlations. (a) For an unknown process of interest, specific separable input states and their corresponding output states (dark gray) can be used to explore the evolution of a state through the unknown process, $\chi_{\rm{expt}}$. Specifically, one can experimentally determine the unknown process by QPT. (b) For a process with no steering capability (denoted as $\chi_{\mathcal{I}}$), all of the separable input states remain unsteerable states~[Eq.~(\ref{eq:5.11})].
Under this scenario, Alice’s output states are considered untrusted (checker pattern). A process having the capability to generate steering is thus a process that cannot be described by $\chi_{\mathcal{I}}$. (c) For a process with no ability to generate Bell nonlocality (denoted as $\chi_{\mathcal{U}}$), the separable input states remain Bell-local states~[Eq.~(\ref{eq:5.12})].
Under this scenario, both output states are considered untrusted (checker pattern). A process having the ability to generate Bell nonlocality is thus a process that cannot be described by $\chi_{\mathcal{U}}$.}
\label{concept}
\end{figure*}

These three quantum correlations are involved in many information processing tasks. Therefore, the processes of generating, preserving, distributing, and applying quantum correlations in QIP are of great interest and importance~\cite{ma2016converting, piani2008no}. Moreover, the figures of merit demonstrated in QIP tasks quantify whether the experimental resources and their processes (including channels) are sufficiently qualified for such quantities. These quantities can thus be regarded as essential for developing useful benchmarks across platforms.

Considerable efforts have been made in determining and quantifying quantum resources, such as entanglement~\cite{vogel1989determination, leonhardt1997measuring, peres1996separability, Horodecki_1996, PhysRevLett.80.2245, PhysRevA.65.032314}, steerability~\cite{cavalcanti2009experimental, skrzypczyk2014quantifying, piani2015necessary, gallego2015resource}, and Bell nonlocality~\cite{bell1964einstein, clauser1969proposed, brunner2014bell, de2014nonlocality, eberhard1993background, popescu1994quantum, toner2003communication, pironio2003violations, van2005statistical, acin2005optimal, junge2010operator, hall2011relaxed, chaves2012multipartite, fonseca2015measure, chaves2015unifying, ringbauer2016experimental, montina2016information, brask2017bell, gallego2017nonlocality, brito2018quantifying}, in order to evaluate the trustworthiness of QIP.  Furthermore, much work has been done on identifying and quantifying different dynamical quantum processes by, for example, identifying whether a given quantum process can create maximum entangled states from separable states~\cite{campbell2010optimal}, witnessing non-entanglement-breaking quantum channels~\cite{moravvcikova2010entanglement, zhen2020unified, mao2020experimentally}, quantifying the properties of quantum channels according to deductive methods~\cite{hsieh2017quantifying, kuo2019quantum}, and utilizing channel resource theory~\cite{chitambar2019quantum, rosset2018resource, theurer2019quantifying, takagi2019general, yuan2021universal, liu2019resource, gour2019quantify, liu2020operational, hsieh2020resource, saxena2020dynamical, takagi2020application, uola2020quantification}.
Particularly, channel resource theory, which is extended from resource theory for quantum states~\cite{chitambar2019quantum} and constructed as a mathematical structure for channel resource, characterizes whether the resources are generated or preserved in the process of interest, and quantifies the channel resources in a completely positive (CP) and trace-preserving (TP) channel~\cite{gour2019quantify, Liu20202020, hsieh2020resource, Saxena20202020}.

However, there are some challenges to quantify the ability of an experimental process in experimentally feasible ways by channel resource theory, as discussed in Ref.~\cite{chen2021quantifying}.
First, the non-TP experimental processes, e.g. photon fusion process~\cite{Zeilinger19981998, Weinfurter20122012}, can't be  analyzed.
Second, whether an experimental process can be simulated by the classical theory~\cite{hsieh2017quantifying} can't be analyzed because it's not related to the state resources required by channel resource theory, and in which there is no definition of the process characteristic.
In addition, to quantify the capability of the process in some tasks of interest, e.g. the preservation capability, the certain resources such as entangled states are required to prepare.
Finally, the optimization of the ancillary system and the superchannel (which includes interactions between the main system and ancillary systems) for the output state to show the clearest difference between the experimental process and the free superchannels (which describe that resource is nonincreasing through free operation) can't be realized necessarily in the experiment.

Considering the experimental feasibility, we use quantum process capability (QPC) theory~\cite{hsieh2017quantifying,kuo2019quantum} rather than channel resource theory to experimentally quantify the ability of the process in this work. QPC method, which is experimentally feasible, requires the input of just certain separable states to obtain entire knowledge of the experimental process based on the primitive measurement results of the output states, and it is reviewed in the following: 1.~QPC theory classifies processes into capable processes, which show the quantum-mechanical effect on a system prescribed by the specification, and incapable processes, which are unable to satisfy the specification at all. Different from channel resource theory, QPC method follows the quantum operations formalism to characterize the experimental process by quantum process tomography (QPT)~\cite{nielsen2002quantum, ChuangPrescription} and quantifies QPC of it. 2.~We can utilize QPC theory to analyze not only CPTP processes but also non-TP CP processes~\cite{nielsen2002quantum, kuo2019quantum}.
Furthermore, QPC theory describes and defines the capability of the whole process to cause quantum-mechanical effects on physical systems as a process characteristic. 3.~QPC theory can be used to characterize the process and quantify its capability for characterizing different QIP tasks.
For example, the work done by Hsieh \textit{et~al.}~\cite{hsieh2017quantifying} and Chen \textit{et~al.}~\cite{chia} showed that the analysis of quantum characteristics from a process perspective is more comprehensive than that from a state perspective. 

Nonetheless, despite the progress made in the studies above in understanding quantum processes, it still remains unclear as to how best to examine the experimental capabilities of creating quantum correlations such as EPR steering and Bell nonlocality. As described above, these resources play a central role in QIP applications. Moreover, they have immediate practical use in quantum computers. However, when it comes to the practical implementation of quantum computers, environmental disturbances or unexpected experimental conditions can bring the output system into compliance with the laws of classical physics. In the worst case, this may cause the calculations to deviate enormously from the expected results~\cite{preskill2018quantum, blais2007quantum, wendin2017quantum, li2019tackling, bharti2021noisy}.

To finally determine and quantify the quantum properties of quantum processes that generate quantum correlations, a fundamental question arises as to the extent to which non-local processes that generate quantum correlations can be accomplished using classical methods of mimicry. The answer has profound implications for how faithfully the QIP task can be implemented and immediately employed for cross-platform benchmark development on NISQ devices. 

To answer this question, we investigate herein the problem of quantifying the generating processes of quantum correlations through the use of tomography tools and numerical methods. To narrow the scope of the study, we focus particularly on just two quantum correlation resources, namely EPR steering and Bell nonlocality. We classify the output states according to different trust situations for both scenarios.
For a process with no capability (ability) to generate EPR steering (Bell nonlocality), separable input states remain unsteerable (Bell-local), and Alice's (both) output states are considered untrusted. See Fig.~\ref{concept}. We show that the proposed approach provides a viable means of quantifying QIP tasks and can be used as the basis for benchmark development on various real-world NISQ devices, including the IBM Q Experience~\cite{ibmq} and AWS Amazon Braket~\cite{aws}.

The remainder of this paper is organized as follows. Section~\ref{sec:quantum correlations} reviews the nonclassical properties of EPR steering and Bell nonlocality, respectively. Section~\ref{sec:cpm} describes the use of quantum state tomography (QST) and QPT~\cite{nielsen2002quantum, ChuangPrescription} to perform the characterization of two-qubit experimental processes. Sections~\ref{sec:sgp} and~\ref{sec:bgp} introduce the steering generating process and Bell nonlocality generating process, respectively. Section~\ref{sec:MQIGPQC} presents several quantifiers, identifiers, and fidelity criteria for evaluating the capability of a process to generate EPR steering and Bell nonlocality quantum correlations. Section~\ref{sec:QOQCGP} presents the quantification and identification results obtained for the quantum correlation generation capability of a two-qubit controlled-phase (CPHASE) shift gate implemented on real-world NISQ devices. Finally, Section~\ref{sec:cao} provides some brief concluding remarks and indicates the intended direction of future research.

\section{\label{sec:quantum correlations}Quantum correlations}

\subsection{EPR steering}

We begin by describing two-qubit states with and without EPR steering, respectively. In identifying the quantum correlations of a composite system, different quantum correlations are associated with different levels of trust in the measurement device. In the case where only Bob's local measurement device is trusted, the violation of steering inequality can be used to prove the existence of EPR steering. When seeking to identify EPR steering, the aim is to determine whether a local measurement device held by Alice can steer the system of another local measurement device held by Bob to a particular state.

Wiseman \textit{et al.} proved that EPR steering exists in a shared pair if, and only if, the correlations between the measurement result of Bob and that announced by Alice cannot be described by the local hidden state (LHS) model~\cite{wiseman2007steering}.
According to this model, the unnormalized state corresponding to the state of Bob's system conditioned on Alice's measurement, ${\rho^{(\rm{B})}_{l}}$, is described as
\begin{equation}\label{neweq:2.13}
P({v^{(\rm{A})}_{k}}){\rho^{(\rm{B})}_{l}}=\sum_{\mu}P({v^{(\rm{A})}_{k}}|{\lambda_\mu})P(\lambda_\mu)\rho^{(\rm{B})}_{\lambda_\mu} \ \ \ \ \ \ \ \forall k, l,
\end{equation}
where
$v^{(\rm{A})}_{k}$ denotes the measurement outcome of Alice's $k$th measurement $V^{(\rm{A})}_{k}$ corresponding to the observable $\hat{V}^{(\rm{A})}_{k}$ with a probability $P({v^{(\rm{A})}_{k}})$,
$\lambda_\mu$ represents Alice's local hidden variable,
$P({v^{(\rm{A})}_{k}}|{\lambda_\mu})$ is the probability of finding $v^{(\rm{A})}_{k}$ conditioned on ${\lambda_\mu}$,
and $\rho^{(B)}_{\lambda_\mu}$ is Bob's local hidden state corresponding to $\lambda_\mu$.
Given Alice's $k$th measurement $V^{(\rm{A})}_{k}$ and outcome $v^{(\rm{A})}_{k}$, any assemblage $\{P({v^{(\rm{A})}_{k}}){\rho^{(\rm{B})}_{l}}\}_{kl}$ of unnormalized states produced in Bob's system due to Alice's measurement which cannot be represented in the form of Eq.~(\ref{neweq:2.13}) is said to be steerable.
Considering $P(v^{(\rm{B})}_{l})=\text{tr}(\ket{v^{(\rm{B})}_{l}}\!\!\bra{v^{(\rm{B})}_{l}\!} {\rho^{(B)}_{l}})$, where $v^{(\rm{B})}_{l}$ denotes the measurement outcome of Bob's $l$th observable $\hat{V}^{(\rm{B})}_{l}$ and $\ket{v^{(\rm{B})}_{l}}\!\!\bra{v^{(\rm{B})}_{l}}$ is the eigenstate corresponding to eigenvalue $v^{(\rm{B})}_{l}$ of the observable $\hat{V}^{(\rm{B})}_{l}$, the LHS model describes the correlation of Alice's and Bob's measurement outcomes as the probability in the following,
\begin{equation}\label{neweq2}
P(v^{(\rm{A})}_{k}, v^{(\rm{B})}_{l})=\sum_{\mu}P({\lambda_\mu})P({v^{(\rm{A})}_{k}}|{\lambda_\mu})\text{tr}(\ket{v^{(\rm{B})}_{l}}\!\!\bra{v^{(\rm{B})}_{l}\!} {\rho^{(B)}_{\lambda_\mu}}),
\end{equation}
where $v^{(\rm{A})}_{k}, v^{(\rm{B})}_{l}\in\pm{1}$ for $k,~l=1,2,3$.

Previous works have adopted various approaches for certifying the existence of EPR steering, including using the steerable weight \cite{skrzypczyk2014quantifying} or robustness of steering \cite{piani2015necessary} to quantify the steering of two-qubit states. 

\subsection{Bell nonlocality}
We next describe the two-qubit state with and without Bell nonlocality~\cite{bell1964einstein}, respectively. Compared to the EPR steering case described above, in which only Alice's measurement device is untrusted, the Bell nonlocality case considers the situation in which neither of the local measurement devices can be trusted.

When both local measurement devices are untrusted, the inability of the local hidden variables (LHV) model~\cite{brunner2014bell} to describe the correlation between the measurements of Alice and Bob, respectively, is seen as evidence of the qubit states with Bell nonlocality.
The LHV model describes the correlation of Alice's and Bob's measurement outcomes as the probability in the following,
\begin{equation}\label{neweq3}
P(v^{(\rm{A})}_{k}, v^{(\rm{B})}_{l})=\sum_{\mu}P({\lambda_\mu})P({v^{(\rm{A})}_{k}}|{\lambda_\mu})P({v^{(\rm{B})}_{l}}|{\lambda_\mu}),
\end{equation}
where $v^{(\rm{A})}_{k}$ and $v^{(\rm{B})}_{l}$ denote Alice's and Bob's measurement outcomes of their $k$th and $l$th measurements $V^{(\rm{A})}_{k}$ and $V^{(\rm{B})}_{l}$ corresponding to the observables $\hat{V}^{(\rm{A})}_{k}$ and $\hat{V}^{(\rm{B})}_{l}$, respectively, and $v^{(\rm{A})}_{k}, v^{(\rm{B})}_{l}\in\pm{1}$ for $k,~l=1,2,3$.

In proving the presence of Bell nonlocality, most previous works use Bell inequalities based on Bell tests \cite{bell1964einstein}. However, it is also possible to quantify the Bell nonlocality of states using nonlocal resources~\cite{toner2003communication,pironio2003violations,montina2016information,steiner2000towards,branciard2011quantifying}, statistical strength measures~\cite{van2005statistical,acin2005optimal}, and the tolerance of nonlocal correlations to noise addition~\cite{junge2010operator,brito2018quantifying,kaszlikowski2000violations,acin2002quantum,perez2008unbounded,massar2002nonlocality}.

\section{\label{sec:cpm}Quantum tomography for a two-qubit experimental process}

In the present study, we characterize an experimental process using a QPT algorithm~\cite{nielsen2002quantum, ChuangPrescription}, which takes as its input 36 pure states underlying the two input objects, i.e., $\ket{\boldsymbol\phi_{i_mj_n}}=\ket{\phi_{i_m}}\otimes\ket{\phi_{j_n}}$, where $\ket{\phi_{i_m}}$ and $\ket{\phi_{j_n}}$ for $i,j=1,2,3$ and $m,n=\pm1$ are the eigenstates corresponding to eigenvalues $m$ and $n$ of the observables $\hat{V}_i$ and $\hat{V}_j$, respectively. Each observable set, $\hat{V}_i$ and $\hat{V}_j$, and the identity matrix form an orthonormal set of matrices with respect to the Hilbert-Schmidt inner product. That is, the observables are chosen as $\hat{V}_1=X$, $\hat{V}_2=Y$, and $\hat{V}_3=Z$.
The Pauli matrixes $X$, $Y$, and $Z$ can be represented as the following spectral decompositions:
$X=\ket{+}\!\!\bra{+}-\ket{-}\!\!\bra{-}$, $Y=\ket{R}\!\!\bra{R}-\ket{L}\!\!\bra{L}$, and $Z=\ket{0}\!\!\bra{0}-\ket{1}\!\!\bra{1}$, where 
$\ket{+}$ and $\ket{-}$ are the eigenstates corresponding to eigenvalues $+1$ and $-1$ of the Pauli-$X$ matrix,
$\ket{R}$ and $\ket{L}$ are the eigenstates corresponding to eigenvalues $+1$ and $-1$ of the Pauli-$Y$ matrix,
and $\ket{0}$ and $\ket{1}$ are the eigenstates corresponding to eigenvalues $+1$ and $-1$ of the Pauli-$Z$ matrix.
Following the description above, we define $\ket{\phi_{1_{1}}} = \ket{+}$, $\ket{\phi_{1_{-1}}} = \ket{-}$, $\ket{\phi_{2_{1}}} = \ket{R}$, $\ket{\phi_{2_{-1}}} = \ket{L}$, $\ket{\phi_{3_{1}}} = \ket{0}$, and $\ket{\phi_{3_{-1}}} = \ket{1}$.
Thus, the $36$ input states can be represented as $\ket{\boldsymbol\phi_{i_mj_n}}\!\!\bra{\boldsymbol\phi_{i_mj_n}}=\ket{\phi_{i_m}}\!\!\bra{\phi_{i_m}}\otimes\ket{\phi_{j_n}}\!\!\bra{\phi_{j_n}}$ for $i,j=1,2,3$ and $m,n=\pm1$.
We assume that the input states are quantum states due to the well quality of the state preparation process in the superconducting computer~\cite{kjaergaard2020superconducting}, and we aim to examine the capability of a process to generate quantum correlations based only upon an inspection of the output states.

Through the QPT algorithm, a positive Hermitian matrix can be used to fully characterize the scenario of a physical process acting on a system. For convenience, the Hermitian matrix is referred to hereinafter simply as the process matrix, $\chi_{\rm{expt}}$. The evolution (operation) of the system from an initial state, $\ket{\boldsymbol\phi_{i_mj_n}}$, to an output state, $\rho_{{\rm{out}}|i_mj_n}$, can then be specified in terms of the process matrix, $\chi_{\rm{expt}}$~\cite{nielsen2002quantum}. Here we denote such operation relating the input and output states by:
\begin{equation}
\label{eq:1}
\chi_{\rm{expt}}(\ket{\boldsymbol\phi_{i_mj_n}}\!\!\bra{\boldsymbol\phi_{i_mj_n}})=\rho_{{\rm{out}}|i_mj_n},
\end{equation}
as shown in Fig.~\ref{concept}(a). From QST, the density operator of the output system conditioned on a specific input state $\ket{\phi_{i_m}}\otimes\ket{\phi_{j_n}}$ can be obtained as
\begin{equation}
\begin{aligned}
\label{eq:2}
\rho_{{\rm{out}}|i_mj_n}=&\frac{1}{4}(\hat{I}\otimes\hat{I}\\
%--------------------
&+\!\!\sum^{3}_{k,l=1}\sum_{v^{(\!\rm{A}\!)}_{k},v^{(\!\rm{B}\!)}_{l}=\pm{1}}\!\!\!
v^{(\!\rm{A}\!)}_{k}\!v^{(\!\rm{B}\!)}_{l}\!P_{i_mj_n}\!(\!v^{(\!\rm{A}\!)}_{k}\!\!,\!v^{(\!\rm{B}\!)}_{l}\!)
\hat{V}^{(\!\rm{A}\!)}_{k}\!\otimes\!\hat{V}^{(\!\rm{B}\!)}_{l}\\
%--------------------
&+\sum^{3}_{k=1}\sum_{v^{(\rm{A})}_{k}=\pm{1}}
v^{(\rm{A})}_{k}P_{i_m}(v^{(\rm{A})}_{k})
\hat{V}^{(\rm{A})}_{k}\otimes\hat{I}\\
%--------------------
&+\sum^{3}_{l=1}\sum_{v^{(\rm{B})}_{l}=\pm{1}}
v^{(\rm{B})}_{l}P_{j_n}(v^{(\rm{B})}_{l})
\hat{I}\otimes\hat{V}^{(\rm{B})}_{l}
),
\end{aligned}
\end{equation}
where $\hat{I}$ is the identity operator, and $P_{i_mj_n}(v^{(\rm{A})}_{k},v^{(\rm{B})}_{l})$ is the joint probability of obtaining the measurement result $v^{(\rm{A})}_{k},v^{(\rm{B})}_{l}\in\pm{1}$ for $k,~l=1,2,3$, conditioned on the specific input $\ket{\boldsymbol\phi_{i_mj_n}}$ of two different subsystems (Alice and Bob, respectively). The input states $\ket{\phi_{i_m}}\otimes\ket{\phi_{j_n}}$ are chosen as the eigenstates of three Pauli matrices, i.e., the Pauli-$X$ matrix, Pauli-$Y$ matrix, and Pauli-$Z$ matrix, for $\hat{V}^{(\rm{A})}_{k}$ and $\hat{V}^{(\rm{B})}_{l}$. Given the corresponding output states, $\rho_{{\rm{out}}|i_mj_n}$, the process matrix can be constructed through the QPT algorithm. See Appendix~\ref{app:qptpm} for details.

\section{\label{sec:sgp}Steering generating process}
In Sec.~\ref{sec:quantum correlations}, we described the concept of two-qubit states with and without EPR steering. In Sec.~\ref{sec:cpm}, we used the positive Hermitian matrix in the QPT algorithm (i.e., the so-called process matrix, $\chi_{\rm{expt}}$) to describe the mapping between the input and output states of a process. This section combines these two concepts to illustrate the steering generation capability of an experimental process. To quantify the steering generation capability of the process, we classify the two-qubit process as either steering generating capable or steering generating incapable using QPC theory \cite{hsieh2017quantifying,kuo2019quantum}.

\textbf{Definition. Incapable process and capable process for steering generating capability.} A process is said to be steering generating incapable, denoted as $\chi_{\mathcal{I}}$, if all the separable input states remain as unsteerable states after the process. Conversely, a process is said to be steering generating capable if the process cannot be described by $\chi_{\mathcal{I}}$ at all.

Since an incapable process $\chi_{\mathcal{I}}$ cannot generate steering from separable states, the output states must be unsteerable states. In the case considered herein, we perform QST on the output states in Eq.~(\ref{eq:2}) according to the QPT algorithm, and thus the measurements we choose are the three Pauli matrices (i.e., Pauli-$X$, Pauli-$Y$, and Pauli-$Z$).

According to the definition of incapable processes, $\chi_{\mathcal{I}}$, the output states in Eq.~(\ref{eq:2}) are unsteerable and can thus be described by the LHS model. For an unsteerable state, the classical state $\textbf{v}^{(\rm{A})}_{\mu}$ can be considered as an object with properties satisfying the assumption of classical realism~\cite{einstein1935can}, i.e., $\textbf{v}^{(\rm{A})}_{\mu}=(\text{v}^{(\rm{A})}_{1},\text{v}^{(\rm{A})}_{2},\text{v}^{(\rm{A})}_{3})$, where $\text{v}^{(\rm{A})}_{1},\text{v}^{(\rm{A})}_{2},\text{v}^{(\rm{A})}_{3} \in ~\{+1,-1\}$ [see Fig.~\ref{concept}(b)]. Thus, the probabilities $P_{i_mj_n}(v^{(\rm{A})}_{k},v^{(\rm{B})}_{l})$ of the output states in Eq.~(\ref{eq:2}) for unsteerable states can be derived from Eq.~(\ref{neweq2}) as
\begin{equation}
P_{i_mj_n}\!(\!v^{(\rm{A})}_{k}\!\!\!,\!v^{(\rm{B})}_{l}\!)\!\!=\!\!\!
\sum_{\mu}\!\!P(\!\textbf{v}^{(\rm{A})}_{\mu}\!)P(\!v^{(\rm{A})}_{k}\!|\textbf{v}^{(\rm{A})}_{\mu}\!)\rm{tr}(\ket{\textit{v}^{(\rm{B})}_\textit{l}}\!\!\bra{\textit{v}^{(\rm{B})}_\textit{l}\!}\!\rho^{(\rm{B})}_{\mu,\textit{i}_\textit{m}\textit{j}_\textit{n}}\!),\label{eq:5.11}
\end{equation}
where
$P_{i_mj_n}(v^{(\rm{A})}_{k}|\textbf{v}^{(\rm{A})}_{\mu})$ are the conditional probabilities; $\rho^{(\rm{B})}_{\mu,i_mj_n}$ is the state held by Bob for the specific input $\ket{\boldsymbol\phi_{i_mj_n}}$; and Alice’s measurement results are decided by $\textbf{v}^{(\rm{A})}_{\mu}$. In accordance with the QPT algorithm in Eq.~(\ref{eq:2}), the observables for the QST procedure for $\rm{Alice}$'s particle are chosen as $\hat{V}^{(\rm{A})}_{1}=X, \hat{V}^{(\rm{A})}_{2}=Y, \hat{V}^{(\rm{A})}_{3}=Z$ while those for Bob’s particle are chosen as $\hat{V}^{(\rm{B})}_{1}=X, \hat{V}^{(\rm{B})}_{2}=Y, \hat{V}^{(\rm{B})}_{3}=Z$. From the classical description in Eq.~(\ref{eq:5.11}), we can replace the joint probability $P_{i_mj_n}(v^{(\rm{A})}_{k},v^{(\rm{B})}_{l})$ in Eq.~(\ref{eq:2}) to construct an incapable process, $\chi_{\mathcal{I}}$, for which all the output states are unsteerable states.
To concretely quantify the ability of an experimental process to generate steering, we propose in this study two quantifiers and a fidelity criterion, as described in Sec.~\ref{sec:MQIGPQC}.

\section{\label{sec:bgp}Bell nonlocality generating process}

This section illustrates the Bell nonlocality generating ability of a process by combining the concept of Bell nonlocality (see Sec.~\ref{sec:quantum correlations}) and the QPT algorithm (see Sec.~\ref{sec:cpm}). In order to identify the Bell nonlocal generating capability of an experimental process, we classify the two-qubit process as either Bell nonlocality generating able or Bell nonlocality generating unable using a similar concept of QPC theory \cite{hsieh2017quantifying,kuo2019quantum}.
  
\textbf{Definition. Unable and able processes for Bell nonlocality generating.} A process is said to be an unable process that cannot generate Bell nonlocality, denoted as $\chi_{\mathcal{U}}$, if all the input separable states remain as Bell-local states following the process. A process is then said to be an able process that can generate Bell nonlocality if it cannot be described by $\chi_{\mathcal{U}}$ at all.

Since an unable process $\chi_{\mathcal{U}}$ cannot generate Bell nonlocality from separable states, the output states must be Bell-local states of which the measurement results can be explained by the LHV model.

% [See Fig.~\ref{fig_expt}(c)]
The LHV model can explain the correlations between the subsystems belonging to $\rm{Alice}$ and $\rm{Bob}$, respectively. In that case, the output states in Eq.~(\ref{eq:2}) can be regarded as a physical object with properties satisfying the assumption of classical realism~\cite{einstein1935can}. The classical states of output systems satisfy the assumption of realism and can be represented by the realistic sets $(\textbf{v}^{(\rm{A})}_{\zeta},\textbf{v}^{(\rm{B})}_{\eta})=(\text{v}_{1}^{(\rm{A})},\text{v}_{2}^{(\rm{A})},\text{v}_{3}^{(\rm{A})},\text{v}_{1}^{(\rm{B})},\text{v}_{2}^{(\rm{B})},\text{v}_{3}^{(\rm{B})})$, where $\text{v}_{1}^{(\rm{A})},\text{v}_{2}^{(\rm{A})},\text{v}_{3}^{(\rm{A})},\text{v}_{1}^{(\rm{B})},\text{v}_{2}^{(\rm{B})},\text{v}_{3}^{(\rm{B})}\in~\{+1,-1\}$ represent the possible measurement outcomes of the $\zeta$th and $\eta$th physical properties of the classical object [see Fig.~\ref{concept}(c)]. Thus, the probabilities $P_{i_mj_n}(v^{(\rm{A})}_{k},v^{(\rm{B})}_{l})$ of the output states in Eq.~(\ref{eq:2}) for Bell-local states can be derived from Eq.~(\ref{neweq3}) as
\begin{equation}
P_{\!i_mj_n\!}\!(\!v\!^{(\rm{A})}_{k}\!\!\!,\!v\!^{(\rm{B})}_{l}\!)\!\!=\!\!\!\sum_{\zeta,\eta}\!\!P_{\!i_mj_n\!}\!(\!\textbf{v}\!^{(\rm{A})}_{\zeta}\!\!,\!\textbf{v}\!^{(\rm{B})}_{\eta}\!)P_{\!i_mj_n\!}\!(\!v\!^{(\rm{A})}_{k}\!|\!\textbf{v}\!^{(\rm{A})}_{\zeta}\!)P_{\!i_mj_n\!}\!(\!v\!^{(\rm{B})}_{l}\!|\!\textbf{v}\!^{(\rm{B})}_{\eta}\!),\label{eq:5.12}
\end{equation}
where
%$\textbf{v}^{(\rm{A})}_{\zeta}$ and $\textbf{v}^{(\rm{B})}_{\eta}$ satisfy the assumption of classical realism, and 
$P_{i_mj_n}(v^{(\rm{A})}_{k}|\textbf{v}^{(\rm{A})}_{\zeta})$ and $P_{i_mj_n}(v^{(\rm{B})}_{l}|\textbf{v}^{(\rm{B})}_{\eta})$ are the conditional probabilities. With the classical description given in Eq.~(\ref{eq:5.12}), we can replace the joint probability $P_{i_mj_n}(v^{(\rm{A})}_{k},v^{(\rm{B})}_{l})$ in Eq.~(\ref{eq:2}) to construct an unable process, $\chi_{\mathcal{U}}$, for which all the output states are Bell-local states.
To concretely identify the ability of an experimental process to generate Bell nonlocality, we propose herein two identifiers and a fidelity criterion, as described in Sec.~\ref{sec:MQIGPQC}.

It is important to note that in constructing an unable process through QPT in Eq.~(\ref{eq:2}), the observables for the QST procedure for $\rm{Alice}$'s particle are selected as $\hat{V}^{(\rm{A})}_{1}=X, \hat{V}^{(\rm{A})}_{2}=Y, \hat{V}^{(\rm{A})}_{3}=Z$ while those for $\rm{Bob}$'s particle must be selected as $\hat{V}^{(\rm{B})}_{1}=U_RX{U_R}^{\dag}, \hat{V}^{(\rm{B})}_{2}=U_RY{U_R}^{\dag}, \hat{V}^{(\rm{B})}_{3}=U_RZ{U_R}^{\dag}$~\cite{Discriminating2020}, where $U_R$ is an arbitrary unitary transformation, i.e.,
\begin{equation}
U_R(\phi,\theta)=\left[ \begin{matrix}
    e^{-i\frac{\phi}{2}}\cos(\frac{\theta}{2}) & e^{-i\frac{\phi}{2}}\sin(\frac{\theta}{2}) \\
    -e^{i\frac{\phi}{2}}\sin(\frac{\theta}{2}) & e^{i\frac{\phi}{2}}\cos(\frac{\theta}{2})
    \end{matrix}
\right].\label{S6}
\end{equation} 
The unitary transformation for the observables $\hat{V}^{B}_{j}$ should be chosen as $U_R(0,\pi/4)$ to maximize the difference between the target process and $\chi_{\mathcal{U}}$. 
Note that we show how to build this unitary transformation in real-world quantum circuits in Sec.~\ref{sec:ecpg}.

\section{\label{sec:MQIGPQC}Methods for quantifying and identifying generating processes of quantum correlations}

This section introduces two quantifiers, two identifiers, and two fidelity criteria for evaluating the capability of an experimental process to generate EPR steering and Bell nonlocality, respectively.

We introduced the definitions of capable and able processes in Sec.~\ref{sec:sgp} and~\ref{sec:bgp}, respectively. We commence this section by constructing a faithful measure, denoted as $C(\chi_{\rm{expt}})$, for a given experimental process matrix $\chi_{\rm{expt}}$, which can be obtained by preparing specific separable states and local measurements in a real-world experiment. We then present two measures (identifiers) for quantifying (identifying) generating processes of quantum correlations, namely (1) the quantum correlation generating composition (i.e., the steering generating composition $\alpha_{\rm{steer}}$ and Bell nonlocality generating composition $\alpha_{\rm{Bell}}$) and (2) the quantum correlation generating robustness (i.e., the steering generating robustness $\beta_{\rm{steer}}$ and Bell nonlocality generating robustness $\beta_{\rm{Bell}}$). Finally, we propose two fidelity criteria for identifying the quantum correlation generating capability of $\chi_{\rm{expt}}$.

Two measures $C(\chi_{\rm{expt}})$ for faithfully quantifying the capability of steering generating processes, i.e., $\alpha_{\rm{steer}}$ and $\beta_{\rm{steer}}$, should satisfy the following three proper measure conditions~\cite{gour2019quantify,liu2020operational,hsieh2020resource}:\\
\textbf{(MP1) Faithfulness:} $C(\chi)=0$ if, and only if, $\chi$ is an incapable process;\\
\textbf{(MP2) Monotonicity:} $C(\chi\circ\chi_{\mathcal{I}})\leq C(\chi)$, i.e., the measures of steering generating capability of a process $\chi$ do not increase following extension with an incapable process;\\
\textbf{(MP3) Convexity:} $C(\sum_{n}p_{n}\chi\circ\chi_{\mathcal{I}})\leq\sum_{n}p_{n}C(\chi\circ\chi_{\mathcal{I}})$, i.e., the mixing of processes does not increase the steering generating capability of the resulting process.

By contrast, two identifiers $C_{\rm{Bell}}(\chi_{\rm{expt}})$ for faithfully identify the capability of Bell nonlocality generating processes, i.e., $\alpha_{\rm{Bell}}$ and $\beta_{\rm{Bell}}$, only satisfy the following two measure conditions:\\
\textbf{(MP1) Faithfulness:} $C_{\rm{Bell}}(\chi)=0$ if, and only if, $\chi$ is an unable process;\\
\textbf{(MP3) Convexity:} $C_{\rm{Bell}}(\sum_{n}p_{n}\chi\circ\chi_{\mathcal{U}})\leq\sum_{n}p_{n}C_{\rm{Bell}}(\chi\circ\chi_{\mathcal{U}})$, i.e., the mixing of processes does not increase the Bell nonlocality generating capability of the resulting process.\\
It is important to note that $\alpha_{\rm{Bell}}$ and $\beta_{\rm{Bell}}$ do not satisfy the monotonicity condition \textbf{(MP2)}, but are still useful for identifying the Bell nonlocality generating process.

\subsection{Composition of quantum correlation generating process} 

A process matrix, $\chi_{\rm{expt}}$, can be expressed as a linear combination of capable processes $\chi_{\mathcal{C}}$ and incapable processes $\chi_{\mathcal{I}}$, or able processes $\chi_{\mathcal{A}}$ and unable processes $\chi_{\mathcal{U}}$, using a similar concept of QPC theory~\cite{hsieh2017quantifying,kuo2019quantum}. That is,
\begin{equation}
\label{eq:15}
\chi_{\rm{expt}}=\alpha_{\rm{steer}}\chi_{\mathcal{C}}+(1-\alpha_{\rm{steer}})\chi_{\mathcal{I}},
\end{equation}
\begin{equation}
\label{eq:15}
\chi_{\rm{expt}}=\alpha_{\rm{Bell}}\chi_{\mathcal{A}}+(1-\alpha_{\rm{Bell}})\chi_{\mathcal{U}},
\end{equation}
where $\alpha_{\rm{steer}}, \alpha_{\rm{Bell}}\ge0$. Let $\alpha_{\rm{steer}}$ and $\alpha_{\rm{Bell}}$ be defined as the minimum amounts of processes that can generate quantum correlations $\chi_{\mathcal{C}}$ and $\chi_{\mathcal{A}}$, respectively, and 
%can be found in the $\chi_{\rm{expt}}$.
have values of $0\ \textless\ \alpha_{\rm{steer}} \leq 1$ for a capable process and $0\ \textless\ \alpha_{\rm{Bell}} \leq 1$ for an able process. 
In practical experiments, $\alpha_{\rm{steer}}$ and $\alpha_{\rm{Bell}}$ can be obtained by minimizing the respective quantities via semi-definite programming (SDP) with MATLAB~\cite{yalmip, sdpt, mosek}:
\begin{equation}
\label{eq:17.2}
\alpha_{\rm{steer}}=\mathop{{\rm{\min }}}\limits_{{\tilde{\chi }}_{\mathcal{I}}}\,[1-{\rm{tr}}({\tilde{\chi }}_{\mathcal{I}})],
\end{equation}
\begin{equation}
\label{eq:17.23}
\alpha_{\rm{Bell}}=\mathop{{\rm{\min }}}\limits_{{\tilde{\chi }}_{\mathcal{U}}}\,[1-{\rm{tr}}({\tilde{\chi }}_{\mathcal{U}})],
\end{equation}
where ${\tilde{\chi }}_{\mathcal{I}}=(1-\alpha_{\rm{steer}})\chi_{\mathcal{I}}$, ${\tilde{\chi }}_{\mathcal{C}}=\alpha_{\rm{steer}}\chi_{\mathcal{C}}$, ${\tilde{\chi }}_{\mathcal{U}}=(1-\alpha_{\rm{Bell}})\chi_{\mathcal{U}}$, and ${\tilde{\chi }}_{\mathcal{A}}=\alpha_{\rm{Bell}}\chi_{\mathcal{A}}$ are unnormalized process matrices with $\text{tr}({\tilde{\chi }}_{\mathcal{I}})=1-\alpha_{\rm{steer}}$ , $\text{tr}({\tilde{\chi }}_{\mathcal{C}})=\alpha_{\rm{steer}}$, $\text{tr}({\tilde{\chi }}_{\mathcal{U}})=1-\alpha_{\rm{Bell}}$, and $\text{tr}({\tilde{\chi }}_{\mathcal{A}})=\alpha_{\rm{Bell}}$, respectively.

In calculating $\alpha_{\rm{steer}}$ of a steering generating process, the constraint set for the SDP problem is given as
\begin{equation}
\label{eq:c17}
\begin{aligned}
&\chi_{\mathcal{I}}\!\geq\!0,\ \chi_{\rm{expt}}\!-\!{\tilde{\chi }}_{\mathcal{I}}\!\geq\!0,\ \chi_{\mathcal{I}}(\ket{\!\boldsymbol\phi_{i_mj_n}\!}\!\!\bra{\!\boldsymbol\phi_{i_mj_n}\!})\!\!\geq\!0,\ \rho_{\mu,i_mj_n}^{(\rm{B})}\!\!\geq\!0,\\
&\sum _{m=\pm1}\rho_{{\rm{out}}|i_mj_n}=\sum _{m=\pm1}\rho_{{\rm{out}}|1_mj_n}, \ \ \ \ \forall \mu,i_m,j_n,
\end{aligned}
\end{equation}
where $\ket{\boldsymbol\phi_{i_mj_n}}\!\!\bra{\boldsymbol\phi_{i_mj_n}}$ are the $36$ input states introduced in Sec.~\ref{sec:cpm} and $\rho_{\mu,i_mj_n}^{(B)}$ are the states held by Bob in Eq.~(\ref{eq:5.11}). 
Similarly, when calculating $\alpha_{\rm{Bell}}$ of a Bell nonlocality generating process, the constraint set is given as
\begin{equation}
\label{eq:18}
\begin{aligned}
&\chi_{\mathcal{U}}\geq0, \ \chi_{\rm{expt}}-{\tilde{\chi }}_{\mathcal{U}}\geq0, \\ &\chi_{\mathcal{U}}(\ket{\boldsymbol\phi_{i_mj_n}}\!\!\bra{\boldsymbol\phi_{i_mj_n}})\geq0, \ P_{i_mj_n}(\textbf{v}^{(\rm{A})}_{\zeta},\textbf{v}^{(\rm{B})}_{\eta})\geq0, \\
&\sum _{m=\pm1}\!\!\!\rho_{{\rm{out}}|i_mj_n}\!\!=\!\!\!\!\sum _{m=\pm1}\!\!\!\rho_{{\rm{out}}|1_mj_n},\!\!\!\!\ \sum _{n=\pm1}\!\!\!\rho_{{\rm{out}}|i_mj_n}\!\!=\!\!\!\!\sum _{n=\pm1}\!\!\!\rho_{{\rm{out}}|i_m1_n}, \\
&\sum _{m=\pm1}\sum _{n=\pm1}\!\!\rho_{{\rm{out}}|i_mj_n}=\!\!\sum _{m=\pm1}\sum _{n=\pm1}\!\!\rho_{{\rm{out}}|1_m1_n},\ \forall i_m,j_n,\zeta,\eta,
\end{aligned}
\end{equation}
where $P_{i_mj_n}(\textbf{v}^{(\rm{A})}_{\zeta},\textbf{v}^{(\rm{B})}_{\eta})$ is the joint probability in Eq.~(\ref{eq:5.12}). 

The constraints $\chi_{\mathcal{I}}\geq0$ and $\chi_{\mathcal{U}}\geq0$ in Eqs. (\ref{eq:c17}) and (\ref{eq:18}), respectively, ensure that the incapable process and unable process are CP mapping. 
Similarly, the constraints $\chi_{\rm{expt}}-{\tilde{\chi }}_{\mathcal{I}}\geq0$ and $\chi_{\rm{expt}}-{\tilde{\chi }}_{\mathcal{U}}\geq0$ ensure that the capable process ${\tilde{\chi }}_{\mathcal{C}}$ and able process ${\tilde{\chi}}_{\mathcal{A}}$ are also CP mapping. 
The constraints $\chi_{\mathcal{I}}(\ket{\boldsymbol\phi_{i_mj_n}}\!\!\bra{\boldsymbol\phi_{i_mj_n}})\geq0$ and $\chi_{\mathcal{U}}(\ket{\boldsymbol\phi_{i_mj_n}}\!\!\bra{\boldsymbol\phi_{i_mj_n}})\geq0$ ensure that all the output states are positive semi-definite for all the input states. 
The fourth constraint in Eq.~(\ref{eq:c17}), i.e., $\rho_{\mu,i_mj_n}^{(\rm{B})}\geq0$, ensures that the density matrix in Eq.~(\ref{eq:5.11}) is positive semi-definite, while the fourth constraint in Eq.~(\ref{eq:18}), i.e., $P_{i_mj_n}(\textbf{v}^{(\rm{A})}_{\zeta},\textbf{v}^{(\rm{B})}_{\eta})\geq0$, ensures that the probability which satisfies the LHV model in Eq.~(\ref{eq:5.12}) is non-negative.

The final constraint in Eq.~(\ref{eq:c17}) ensures that Alice's inputs form an identity matrix $\hat{I}$. The corresponding output states are the same for different decompositions of $\hat{I}$ since the identity matrix $\hat{I}$ can be represented as the sum of the inputs for each basis, i.e., $\hat{I}=\Sigma_{m=\pm1}\ket{\phi_{i_m}}\!\!\bra{\phi_{i_m}}$, $\forall i$. The constraint describes the relations of the outputs in SDP though ensuring that the outputs corresponding to the sum of the inputs for each basis $\sum _{m=\pm1}\rho_{{\rm{out}}|i_mj_n}$, $i=1,2,3$, are equal to the sum in the first basis, $\sum _{m=\pm1}\rho_{{\rm{out}}|1_mj_n}$. Similarly, the remaining constraints in Eq.~(\ref{eq:18}) ensure that when the input states are an identity matrix $\hat{I}$, the corresponding output states are the same for all different decompositions of $\hat{I}$. 

The sixth and seventh constraints in Eq.(\ref{eq:18}) state that the input of Bob's qubit and the input of both qubits are $\hat{I}$, and the outputs $\chi_{\mathcal{U}}(\ket{\phi_{i_m}}\!\!\bra{\phi_{i_m}}\otimes\hat{I})=\sum _{n=\pm1}\rho_{{\rm{out}}|i_m1_n}$ and $\chi_{\mathcal{U}}(\hat{I}\otimes\hat{I})=\sum _{m=\pm1}\sum _{n=\pm1}\rho_{{\rm{out}}|1_m1_n}$, are the same for all decompositions of $\hat{I}$, respectively.

In general, the constraint set determines the number of variables that need to be optimized via SDP. For a two-qubit system, each $\rho_{{\rm{out}}|i_mj_n}$ in the constraint set has the form of a $4\times4$ matrix that contains $16$ variables and is conditioned on $36$ input states. 
To describe the classical dynamics of an incapable process $\chi_{\mathcal{I}}\geq0$, we need $8$ matrices, which corresponds to the number of $\textbf{v}^{(\rm{A})}_{\mu}$. Consequently, a total of $4608$ variables must be solved by SDP. 
Similarly, to describe the classical dynamics of an unable process $\chi_{\mathcal{U}}\geq0$, we need $64$ matrices, which corresponds to the number of $\textbf{v}^{(\rm{A})}_{\zeta}$ and $\textbf{v}^{(\rm{B})}_{\eta}$, and hence $36864$ variables must be solved by SDP.

\subsection{Robustness of quantum correlation generating process}

An incapable steering generating process  or unable Bell nonlocality generating process can be obtained by adding a certain amount of noise into an experimental process, $\chi_{\rm{expt}}$, using a similar concept of QPC theory~\cite{hsieh2017quantifying,kuo2019quantum}, i.e., 
\begin{equation}
\label{eq:19.1}
\frac{\chi {}_{{\rm{expt}}}+\beta_{\rm{steer}} \chi_{\rm{noise}}}{1+\beta_{\rm{steer}} }={\chi}_{\mathcal{I}},
\end{equation}
\begin{equation}
\label{eq:19.2}
\frac{\chi {}_{{\rm{expt}}}+\beta_{\rm{Bell}} \chi_{\rm{noise}}}{1+\beta_{\rm{Bell}} }={\chi}_{\mathcal{U}},
\end{equation}
where $\beta_{\rm{steer}}, \beta_{\rm{Bell}}\ge0$, and $\chi_{\rm{noise}}$ is the noise process. Here, $\beta_{\rm{steer}}$ and $\beta_{\rm{Bell}}$ represent the minimum amounts of noise which must be added to $\chi_{\rm{expt}}$ to turn $\chi_{\rm{expt}}$ into ${\chi }_{\mathcal{I}}$ and ${\chi }_{\mathcal{U}}$, respectively. 
$\beta_{\rm{steer}}$ and $\beta_{\rm{Bell}}$ can be obtained by minimizing ${\chi_{\rm{noise}}}$ via SDP with MATLAB~\cite{yalmip, sdpt, mosek} as follows:
\begin{equation}
\label{eq:17.1}
\beta_{\rm{steer}}=\mathop{{\rm{\min }}}\limits_{{\tilde{\chi }}_{\mathcal{I}}}\,[{\rm{tr}}({\tilde{\chi}}_{\mathcal{I}})-1],
\end{equation}
\begin{equation}
\label{eq:17.2}
\beta_{\rm{Bell}}=\mathop{{\rm{\min }}}\limits_{{\tilde{\chi }}_{\mathcal{U}}}\,[{\rm{tr}}({\tilde{\chi}}_{\mathcal{U}})-1],
\end{equation}
where ${\tilde{\chi }}_{\mathcal{I}}=(1+\beta_{\rm{steer}})\chi_{\mathcal{I}}$ and ${\tilde{\chi }}_{\mathcal{U}}=(1+\beta_{\rm{Bell}})\chi_{\mathcal{U}}$ are unnormalized process matrices with $\text{tr}({\tilde{\chi }}_{\mathcal{I}})=1+\beta_{\rm{steer}}$ and $\text{tr}({\tilde{\chi }}_{\mathcal{U}})=1+\beta_{\rm{Bell}}$, respectively.

In calculating $\beta_{\rm{steer}}$ for a steering generating process, the constraint set is given as
\begin{equation}
\label{eq:17}
\begin{aligned}
&\chi_{\mathcal{I}}\geq0,\ {\tilde{\chi }}_{\mathcal{I}}-\chi_{\rm{expt}}\geq0,\ \text{tr}({\tilde{\chi }}_{\mathcal{I}})\geq1,\\
&\chi_{\mathcal{I}}(\ket{\boldsymbol\phi_{i_mj_n}}\!\!\bra{\boldsymbol\phi_{i_mj_n}})\geq0,\ \rho_{\mu,i_mj_n}^{(\rm{B})} \geq0,\\
&\sum _{m=\pm1}\rho_{{\rm{out}}|i_mj_n}=\sum _{m=\pm1}\rho_{{\rm{out}}|1_mj_n}, \ \ \ \ \forall \mu,i_m,j_n.
\end{aligned}
\end{equation}
Similarly, when calculating $\beta_{\rm{Bell}}$ for a Bell nonlocality generating process, the constraint set is given as
\begin{equation}
\label{eq:c18b}
\begin{aligned}
&\chi_{\mathcal{U}}\geq0,\ {\tilde{\chi }}_{\mathcal{U}}-\chi_{\rm{expt}}\geq0,\ \text{tr}({\tilde{\chi }}_{\mathcal{U}})\geq1, \\
&\chi_{\mathcal{U}}(\ket{\boldsymbol\phi_{i_mj_n}}\!\!\bra{\boldsymbol\phi_{i_mj_n}})\geq0,\ P_{i_mj_n}(\textbf{v}^{(\rm{A})}_{\zeta},\textbf{v}^{(\rm{B})}_{\eta})\geq0, \\
&\sum _{m=\pm1}\!\!\!\rho_{{\rm{out}}|i_mj_n}\!\!=\!\!\!\sum _{m=\pm1}\!\!\!\rho_{{\rm{out}}|1_mj_n},\!\!\sum _{n=\pm1}\!\!\!\rho_{{\rm{out}}|i_mj_n}\!\!=\!\!\!\sum _{n=\pm1}\!\!\!\rho_{{\rm{out}}|i_m1_n}, \\
&\sum _{m=\pm1}\sum _{n=\pm1}\!\rho_{{\rm{out}}|i_mj_n}\!=\!\sum _{m=\pm1}\sum _{n=\pm1}\!\rho_{{\rm{out}}|1_m1_n},\ \forall i_m,j_n,\zeta,\eta.
\end{aligned}
\end{equation}

The constraints ${\tilde{\chi }}_{\mathcal{I}}-\chi_{\rm{expt}}\geq0$ and ${\tilde{\chi }}_{\mathcal{U}}-\chi_{\rm{expt}}\geq0$ ensure that the noise processes ${\tilde{\chi }}_{\rm{noise}}$ in Eqs.~(\ref{eq:19.1}) and~(\ref{eq:19.2}), respectively, are CP mapping. 
Meanwhile, the constraints $\text{tr}({\tilde{\chi }}_{\mathcal{I}})\geq1$ and $\text{tr}({\tilde{\chi }}_{\mathcal{U}})\geq1$ ensure that $\beta_{\rm{steer}}\geq0$ and $\beta_{\rm{Bell}}\geq0$, respectively. 
Note that the other constraints are as described above for the composition measure [Eqs. (\ref{eq:c17}) and (\ref{eq:18})]. Similarly, the numbers of variables to be solved in the SDP optimization tasks in Eqs.~(\ref{eq:17.1}) and~(\ref{eq:17.2}) are also $4608$ and $36864$, respectively.

\subsection{Fidelity criterion for quantum correlations generating process} 

An experimental process, $\chi_{\rm{expt}}$, is created with respect to a target process, $\chi_{\rm{target}}$, and the similarity between them can be assessed using the process fidelity, $F_{\rm{expt}}\equiv \rm{tr}(\chi_{\rm{expt}}\chi_{\rm{target}})$. 
In particular, $F_{\rm{expt}}$ is judged to indicate the capability (ability) of a process to generate steering (Bell nonlocality) if its value goes beyond the best mimicry achieved by incapable processes $\chi_{\mathcal{I}}$ and unable processes $\chi_{\mathcal{U}}$ respectively, i.e.,
\begin{equation}
\label{eq:21.1}
F_{\rm{expt}} \textgreater F_{\mathcal{I}}\equiv\mathop{{\rm{\max }}}\limits_{\chi_{\mathcal{I}}}[\rm{tr}(\chi_{\mathcal{I}}\chi_{\rm{target}})],
\end{equation}
\begin{equation}
\label{eq:21.2}
F_{\rm{expt}} \textgreater F_{\mathcal{U}}\equiv\mathop{{\rm{\max }}}\limits_{\chi_{\mathcal{U}}}[\rm{tr}(\chi_{\mathcal{U}}\chi_{\rm{target}})].
\end{equation}
Eqs.~(\ref{eq:21.1}) and~(\ref{eq:21.2}) show the $\chi_{\rm{expt}}$ cannot be mimicked by any incapable processes and unable processes, respectively. 
The best mimicry achieved by $\chi_{\mathcal{I}}$ and $\chi_{\mathcal{U}}$ can be evaluated by performing the following SDP maximization tasks in MATLAB~\cite{yalmip, sdpt, mosek}:
\begin{equation}
\label{eq:5.22.1}
F_{\mathcal{I}}\equiv\mathop{{\rm{\max }}}\limits_{\chi_{\mathcal{I}}}[\rm{tr}(\chi_{\mathcal{I}}\chi_{\rm{target}})],
\end{equation} 
\begin{equation}
\label{eq:5.22.2}
F_{\mathcal{U}}\equiv\mathop{{\rm{\max }}}\limits_{\chi_{\mathcal{U}}}[\rm{tr}(\chi_{\mathcal{U}}\chi_{\rm{target}})],
\end{equation} 
such that $\rm{tr}(\chi_{\mathcal{I}})=1$ and $\rm{tr}(\chi_{\mathcal{U}})=1$, respectively. 

In calculating the fidelity of a steering generating process, the constraint set is specified as
\begin{equation}
\label{eq:22}
\begin{aligned}
&\chi_{\mathcal{I}}\geq0,\ \chi_{\mathcal{I}}(\ket{\boldsymbol\phi_{i_mj_n}}\!\!\bra{\boldsymbol\phi_{i_mj_n}})\geq0,\ \rho_{\mu,i_mj_n}^{(\rm{B})} \geq0, \\
&\sum _{m=\pm1}\rho_{{\rm{out}}|i_mj_n}=\sum _{m=\pm1}\rho_{{\rm{out}}|1_mj_n}, \ \ \ \ \forall \mu,i_m,j_n.
\end{aligned}
\end{equation}
Similarly, in calculating the fidelity of the Bell nonlocality generating process, the constraint set is given as
\begin{equation}
\label{eq:22}
\begin{aligned}
&\chi_{\mathcal{U}}\geq0,\ \chi_{\mathcal{U}}(\ket{\boldsymbol\phi_{i_mj_n}}\!\!\bra{\boldsymbol\phi_{i_mj_n}})\geq0,\ P_{i_mj_n}(\textbf{v}^{(\rm{A})}_{\zeta},\textbf{v}^{(\rm{B})}_{\eta})\geq0, \\
&\sum _{m=\pm1}\!\!\!\rho_{{\rm{out}}|i_mj_n}\!\!=\!\!\!\sum _{m=\pm1}\!\!\rho_{{\rm{out}}|1_mj_n},\!\!\sum _{n=\pm1}\!\!\!\rho_{{\rm{out}}|i_mj_n}\!\!=\!\!\!\sum _{n=\pm1}\!\!\rho_{{\rm{out}}|i_m1_n}, \\
&\sum _{m=\pm1}\sum _{n=\pm1}\rho_{{\rm{out}}|i_mj_n}\!=\!\sum _{m=\pm1}\sum _{n=\pm1}\rho_{{\rm{out}}|1_m1_n},\ \forall i_m,j_n,\zeta,\eta.
\end{aligned}
\end{equation}
Note that the constraints are all as described above for the composition measure [Eqs. (\ref{eq:c17}) and (\ref{eq:18})].  Note also that the numbers of variables to be solved in the optimization tasks in Eqs.~(\ref{eq:5.22.1}) and~(\ref{eq:5.22.2}) are again $4608$ and $36864$, respectively.

\section{\label{sec:QOQCGP}Quantification and identification of quantum correlation generating processes} 
This section commences by describing two experimental tests for quantum correlation generating processes (a steering generating test and a Bell nonlocality generating test) on circuit-based superconducting quantum computers (Sec.~\ref{sec:ecpg}). The quantification and identification results obtained using the tools described in Sec.~\ref{sec:MQIGPQC} for an ideal simulation, which targets the CPHASE gate, are then presented (Sec.~\ref{sec:sf}). Finally, the results obtained by the proposed quantitative and identifying tools for two real NISQ devices (IBM~Q Experience \cite{ibmq} and Amazon Braket \cite{aws}) and their simulators with and without noise models, respectively, are presented and discussed (Sec.~\ref{sec:QCSQC}). 

\subsection{~\label{sec:ecpg}Implementation of controlled-phase gate on superconducting quantum computer}

In quantum computing, a controlled-phase gate is a two-qubit operation that induces a particular phase of the state of a target qubit depending on the state of the control qubit. The gate is a key element in creating entanglement in superconducting quantum computers and serves as a preliminary logical gate in helping create a graph state \cite{HNN}. 
The two-qubit gate operation can be expressed as follows:
\begin{equation}
\begin{aligned}
\label{CPHASEEEE}
\rm{CPHASE}(\lambda)=\begin{pmatrix}
 1&0&0&0\\
 0&1&0&0\\
 0&0&1&0\\
 0&0&0&e^{i\lambda}
\end{pmatrix},
\end{aligned}
\end{equation}
where $\lambda$ is the CPHASE shift, i.e., the shift of the phase of the target qubit induced by the state of the control qubit. Given this knowledge of the CPHASE gate and the quantitative and identifying tools presented in Sec.~\ref{sec:MQIGPQC}, we present in the following the detailed steps of the QPT procedure~\cite{nielsen2002quantum} for the CPHASE gate on the IBM Q Experience \cite{ibmq} and Amazon Braket \cite{aws} quantum computers. 

\begin{figure}[t]
\includegraphics[width=6cm]{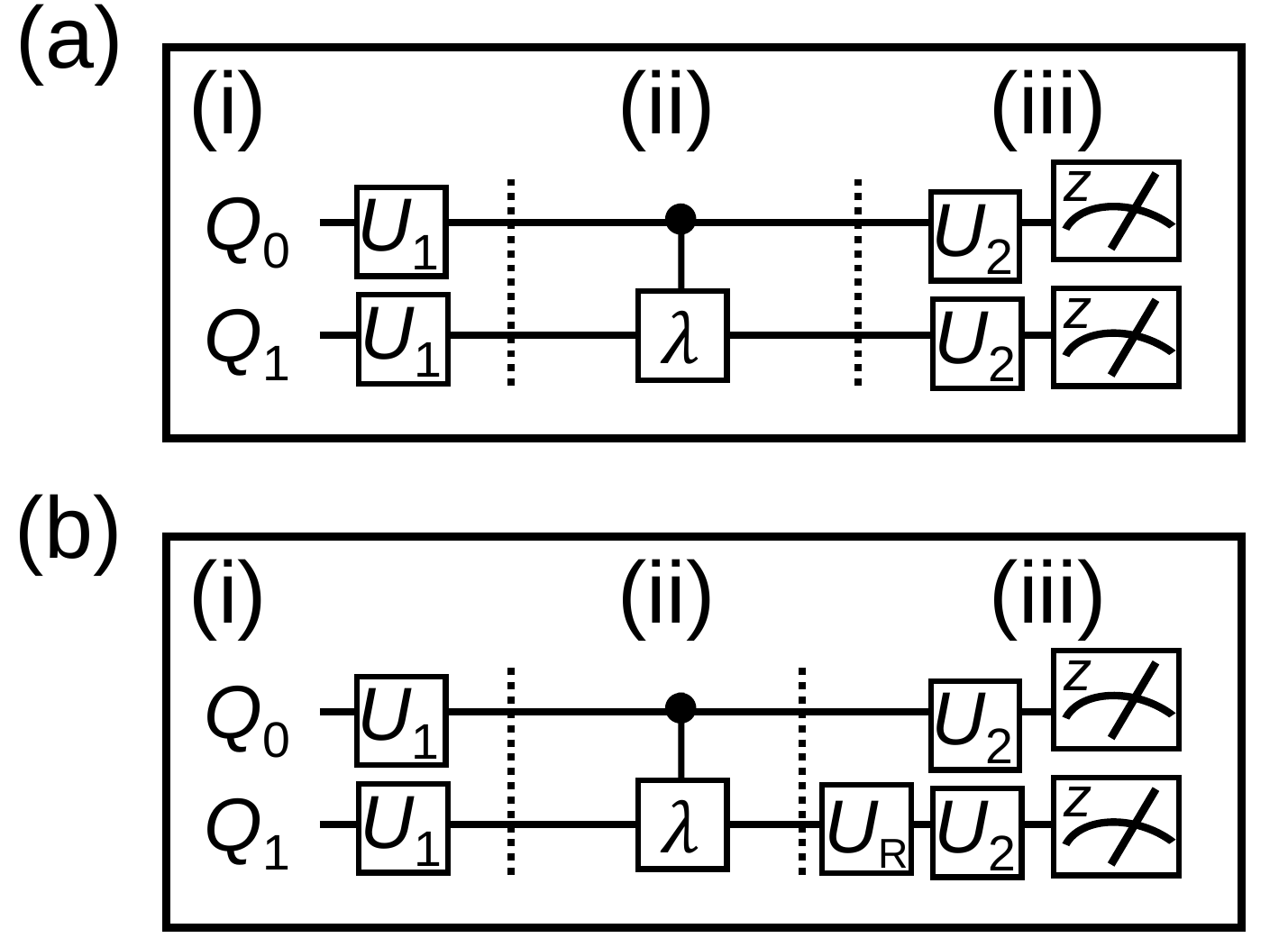}
\caption{\label{fig:schematicPT}Schematic representations of the quantum circuits used to implement the QPT procedure for the steering generating test (a) and Bell nonlocality generating test (b). (i) The notation $Q_{0}$ ($Q_{1}$) indicates logical qubit $0$ ($1$), and $U_1$ is the corresponding unitary transform used to prepare the specific input state. (ii) The CPHASE gate is performed on control qubit $Q_0$ with a CPHASE shift of $\lambda$ on the target qubit $Q_1$. (iii) $U_R$ is a suitable unitary transformation to maximize the difference between the target process and $\chi_{\mathcal{U}}$, and $U_2$ is chosen for the measurements performed on the Pauli-$X$ basis and Pauli-$Y$ basis..
}
\label{qcircuit}
\end{figure}

We first present two tests, namely a steering generating test and a Bell nonlocality generating test for evaluating the quantum correlation generating capability of an experiment process. The quantum circuit implementations of the two tests are shown in Figs.~\ref{fig:schematicPT}(a) and~\ref{fig:schematicPT}(b), respectively. All of the qubits, i.e., $Q_{0}$ and $Q_{1}$, are initially prepared in a state $\ket{\phi_{3_{1}}} = \ket{0}$ (defined in Sec.~\ref{sec:cpm}), and specific quantum states are then prepared as input states by applying the unitary operation $U_1$ [Figs.~\ref{fig:schematicPT}(a)(i) and \ref{fig:schematicPT}(b)(i)]. The input states are then processed by a CPHASE gate with a shift of $\lambda$ [Figs.~\ref{fig:schematicPT}(a)(ii) and \ref{fig:schematicPT}(b)(ii)]. 
Finally, QST~[Eq.~(\ref{eq:2})] is applied to the two output qubits to reconstruct the density matrix, $\rho_{{\rm{out}}|i_mj_n}$, by measuring their states in the Pauli bases \{$X$, $Y$, $Z$\} [Figs.~\ref{fig:schematicPT}(a)(iii) and \ref{fig:schematicPT}(b)(iii)]. Experimentally, the measurements performed on the Pauli-$X$ basis and Pauli-$Y$ basis are implemented by using different transformations $U_2$ followed by measurement on the Pauli-$Z$ basis. In particular, the measurement on the Pauli-$X$ basis is implemented by a Hadamard gate ($H = \ket{+}\!\!\bra{0}+\ket{-}\!\!\bra{1}$)  followed by measurement on the Pauli-$Z$ basis, while the measurement on the Pauli-$Y$ basis is implemented by an adjoint of phase gate ($S^{\dag} = \ket{0}\!\!\bra{0}-i\ket{1}\!\!\bra{1}$) and an $H$ gate followed by measurement on the Pauli-$Z$ basis.

As discussed in Sec.~\ref{sec:bgp}, in the Bell nonlocality generating test, the unitary operation $U_R$ [Fig.~\ref{fig:schematicPT}(b)(iii)] is added to different choices of Bob's observables which generate the most significant difference possible of $F_{\mathcal{U}}$ between the target process and $\chi_{\mathcal{U}}$. Having chosen a suitable unitary operation $U_R(\phi,\theta)$, it is realized in the quantum circuit model by applying the one-qubit gates to the qubit. For example, if $U_R(\phi,\theta)$ is chosen as $\phi=0$ and $\theta=1/4\pi$, it is realized by setting
\begin{equation}
U_R(0,\frac{\pi}{4}) = R_x(\frac{\pi}{2})R_z(\frac{\pi}{4})R_x(-\frac{\pi}{2}),
\end{equation} 
where 
\begin{equation}
R_x(\vartheta)=\left[ \begin{matrix}
    \cos(\frac{\vartheta}{2}) & -i\sin(\frac{\vartheta}{2}) \\
    -i\sin(\frac{\vartheta}{2}) & \cos(\frac{\vartheta}{2})
    \end{matrix}
\right],
R_z(\varphi)=\left[ \begin{matrix}
    e^{-i\frac{\varphi}{2}} & 0 \\
    0 & e^{i\frac{\varphi}{2}}
    \end{matrix}
\right].\label{gate}
\end{equation}

We experimentally determined the process matrix of the two process correlation generating tests on the IBM~Q and Amazon Braket quantum computers (both real devices and simulators). For each run of the quantum circuit, we adjusted the maximum shot setting (i.e., the number of times the qubits of the quantum circuit were measured) in accordance with the limit of the respective device. In particular, we set $1024$ shots for the \textit{Amazon Braket Rigetti Aspen-9} and \textit{Aspen-M-1} devices, $8192$ shots for the \textit{ibmq\_santiago} device, and $81920$ shots for the \textit{Amazon Braket local simulator} device and \textit{ibmq\_qasm\_simulator} device.

\subsection{\label{sec:sf}Quantification and identification of ideal controlled-phase gate}

This section presents the results obtained when applying the quantitative and identifying methods proposed in Sec.~\ref{sec:MQIGPQC} to the steering generating test and Bell nonlocality generating test described above for the case of an ideal CPHASE gate process.

\begin{figure}[t]
\includegraphics[width=8.5cm]{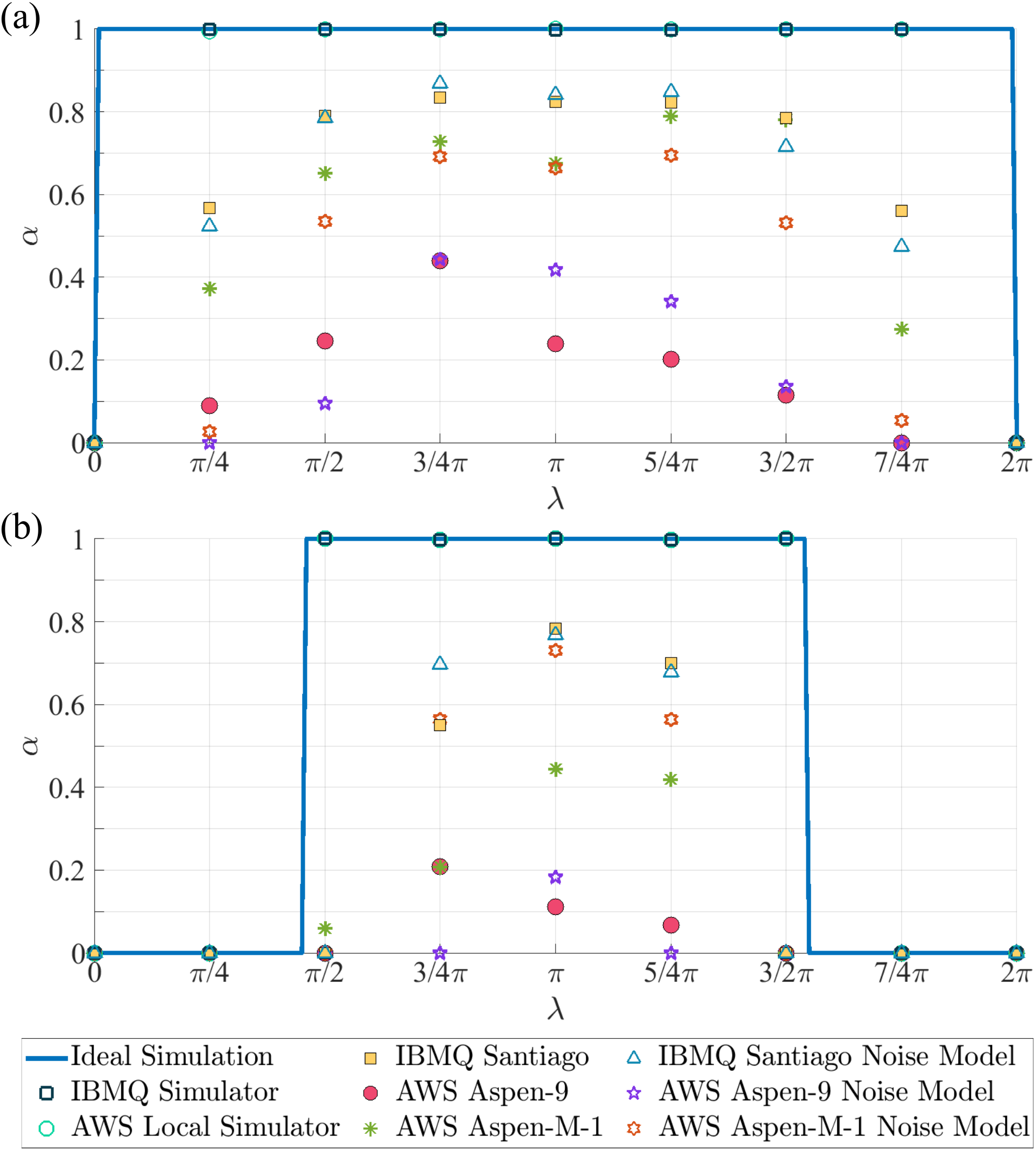}
\caption{\label{fig_sb_a}
Composition results for steering generating test [Fig.~\ref{fig_sb_a}(a)]  and Bell nonlocality generating test [Fig.~\ref{fig_sb_a}(b)]. 
The solid blue line shows the steering (Bell nonlocality) generating composition results from the ideal simulation. The other symbols show the results obtained on the IBM Q Experience and AWS Amazon Braket devices (real and simulator).
}
\label{qcircuit}
\end{figure}

\begin{figure}[t]
\includegraphics[width=8.5cm]{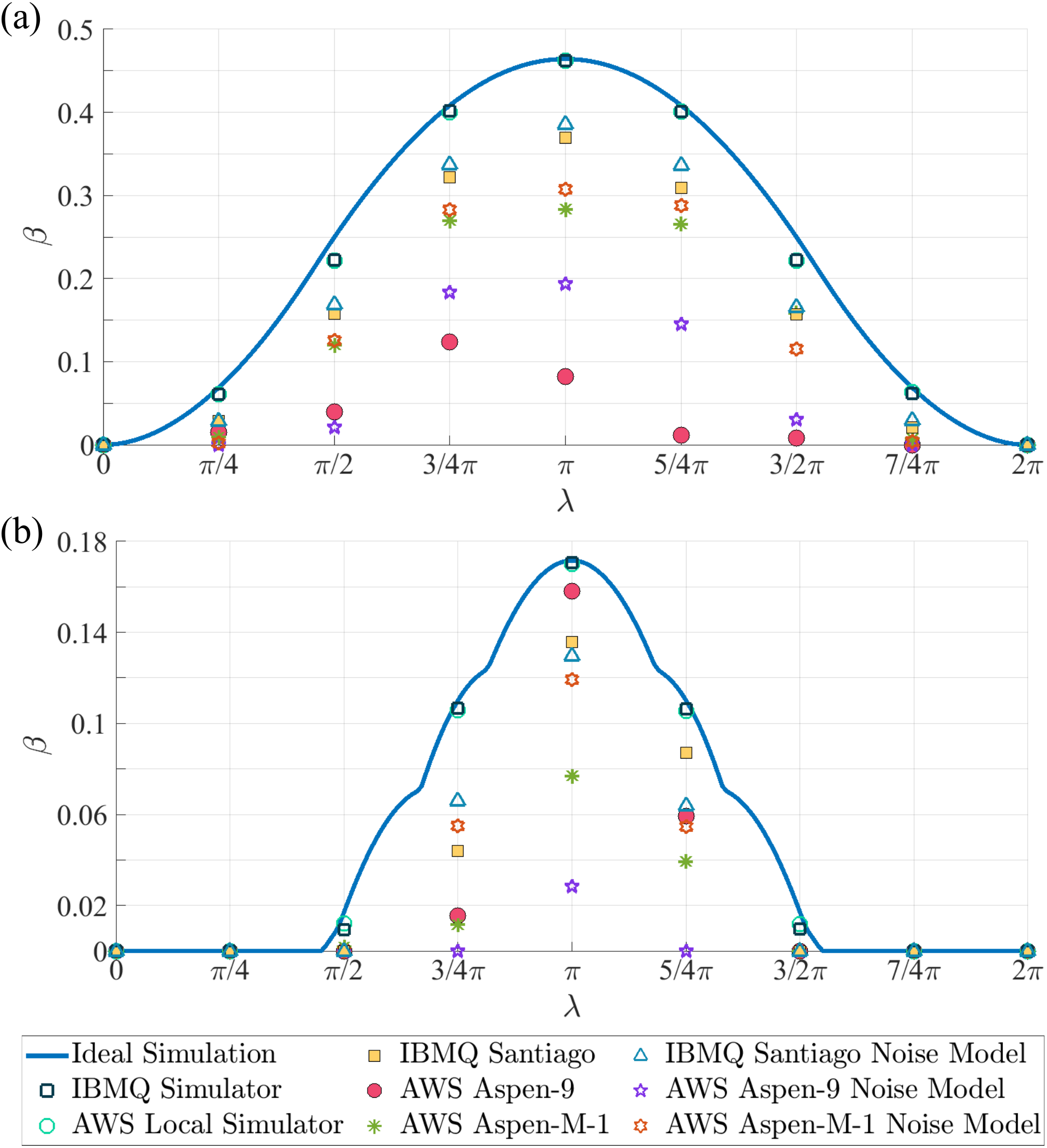}
\caption{\label{fig_sb_b}
Robustness results for steering generating test [Fig.~\ref{fig_sb_b}(a)] and Bell nonlocality generating test [Fig.~\ref{fig_sb_b}(b)]. The solid blue line shows the steering (Bell nonlocality) generating robustness results from the ideal simulation.
The other symbols show the results obtained on the IBM Q Experience and AWS Amazon Braket devices (real and simulator).
} 
\label{qcircuit}
\end{figure}

\begin{figure}[t]
\includegraphics[width=8.75cm]{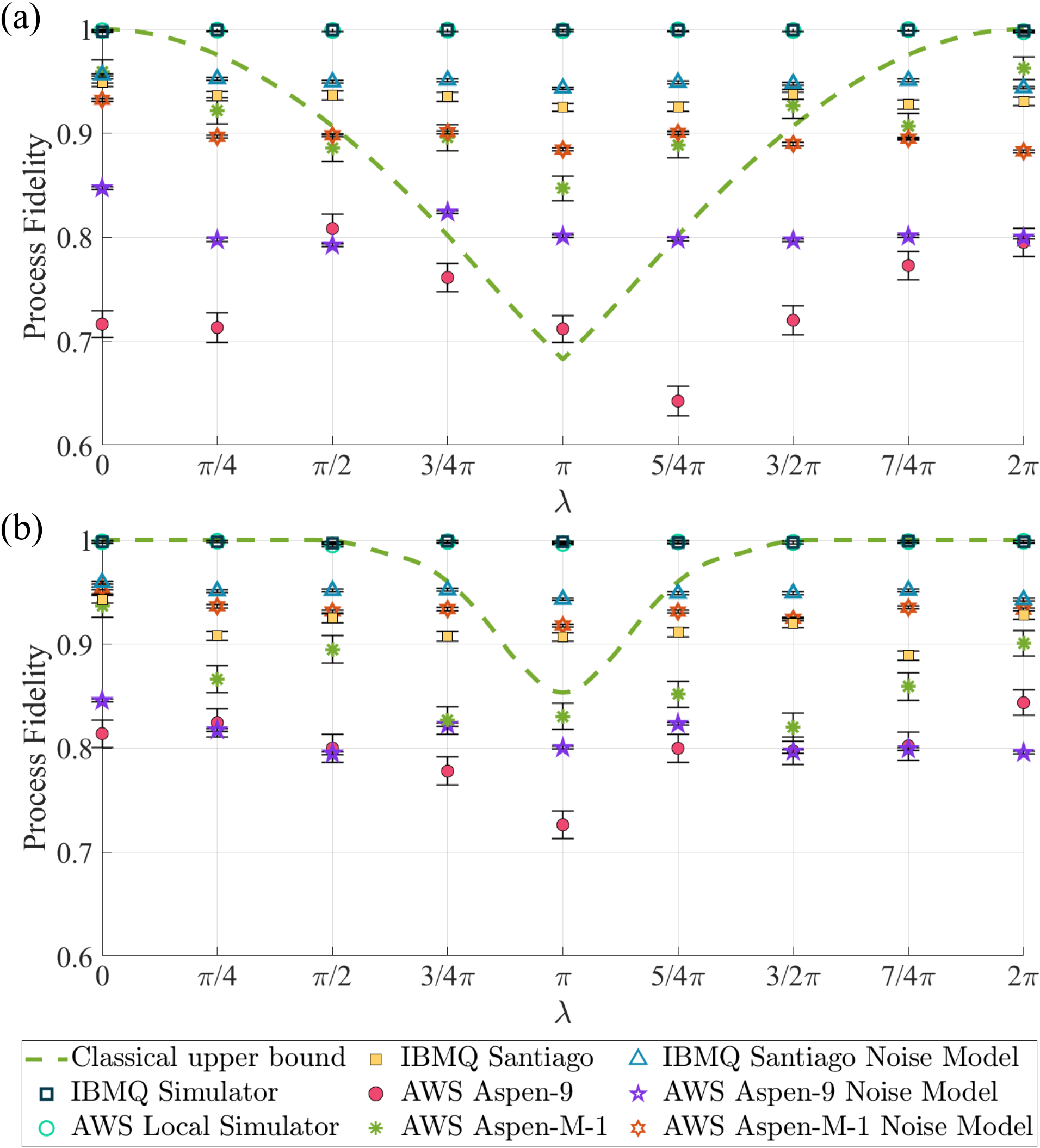}
\caption{\label{fig_sb_f}
Fidelity criterion results for steering generating test [Fig.~\ref{fig_sb_f}(a)] and Bell nonlocality generating test [Fig.~\ref{fig_sb_f}(b)]. The dashed green line shows the classical upper bound of the fidelity criterion for the steering (Bell nonlocality) generating test.
The other symbols show the results obtained on the IBM Q Experience and AWS Amazon Braket devices (real and simulator).
} 
\label{qcircuit}
\end{figure}

In conducting the steering (Bell nonlocality) generating test, we consider an experimental process $\chi_{\rm{expt}}$ that performs an ideal CPHASE gate with a phase shift of $\lambda=\pi$ to be a capable (able) process. 
For such a process, the steering (Bell nonlocality) generating composition $\alpha_{\rm{steer}}$ ($\alpha_{\rm{Bell}}$) has a value of $1$, while the steering (Bell nonlocality) generating robustness $\beta_{\rm{steer}}$ ($\beta_{\rm{Bell}}$) has a value of $0.4641$ ($0.1716$). Figs.~\ref{fig_sb_a}(a) and~\ref{fig_sb_a}(b) show the computed values of $\alpha_{\rm{steer}}$ and $\alpha_{\rm{Bell}}$, respectively, for various values of the CPHASE shift. Figs.~\ref{fig_sb_b}(a) and~\ref{fig_sb_b}(b) show the corresponding results for $\beta_{\rm{steer}}$ and $\beta_{\rm{Bell}}$, respectively. Note that, in both figures, the solid blue lines show the results obtained from the ideal simulation. 
We next consider a process that performs an ideal CPHASE gate with a phase shift of $\lambda=\pi$ as the target process and a process fidelity $F_{\mathcal{I}}$ ($F_{\mathcal{U}}$) of $0.6830$ ($0.8536$). 
The dashed green lines in Figs.~\ref{fig_sb_f}(a) and~\ref{fig_sb_f}(b) show the variation of the fidelity criterion $F_{\mathcal{I}}$ ($F_{\mathcal{U}}$) as the CPHASE shift of the target process changes from $0$ to $2\pi$.

It is seen in Figs.~\ref{fig_sb_a}(b) and~\ref{fig_sb_b}(b) that $\alpha_{\rm{Bell}}$ and $\beta_{\rm{Bell}}$ both start to produce a value at $\lambda=0.46\pi$ and return to zero at $\lambda=1.54\pi$. Meanwhile in Fig.~\ref{fig_sb_f}(b), the fidelity criterion $F_{\mathcal{U}}$ begins to decrease at $\lambda=0.46\pi$, reaches a minimum value of $F_{\mathcal{U}}\sim0.8536$ at $\lambda=\pi$, and then increases once again to a value of around one at $\lambda=1.54\pi$.

\subsection{\label{sec:QCSQC}Quantification and identification of controlled-phase gate on superconducting quantum computer}

The discussions above have described the results obtained from the proposed quantitative and identifying methods for the steering generating test and Bell nonlocality generating test in the ideal simulation of the CPHASE gate.
This section presents the quantification and identification results obtained for an actual CPHASE gate process performed on two superconducting quantum computer systems, namely the IBM Q Experience~\cite{ibmq} and Amazon Braket~\cite{aws}, and their respective simulators.

We selected nine different CPHASE shifts ($\lambda=0$, $1/4\pi$, $1/2\pi$, $3/4\pi$, $\pi$, $5/4\pi$, $3/2\pi$, $7/4\pi$ and $2\pi$) for the CPHASE gate implemented on the superconducting quantum computer system. To perform QPT, we conducted QST on the experimental outputs. The density matrix $\rho_{\rm{out}|\textit{i}_\textit{m}\textit{j}_\textit{n}}$ of the output states was reconstructed using the following 16 specific input states: 
$\{\ket{00}$, $\ket{01}$, $\ket{0+}$, $\ket{0R}$, $\ket{10}$, $\ket{11}$, $\ket{1+}$, $\ket{1R}$, $\ket{+0}$, $\ket{+1}$, $\ket{++}$, $\ket{+R}$, $\ket{R0}$, $\ket{R1}$, $\ket{R+}$, $\ket{RR}\}$ (defined in Sec.~\ref{sec:cpm}).
Having obtained the density matrix, we reconstructed the physical process via QPT. In particular, we experimentally determined the reasonable process matrix of the CPHASE gate using the maximum-likelihood technique~\cite{o2004quantum}.
Finally, we examined the physical process matrix using the composition, robustness, and process fidelity proposed in Sec.~\ref{sec:MQIGPQC}. The corresponding results are presented in Figs.~\ref{fig_sb_a},~\ref{fig_sb_b} and~\ref{fig_sb_f}, respectively.

As described above in Sec.~\ref{sec:sf}, 
the ideal CPHASE gate with a shift of $\lambda=\pi$ possesses the maximum amount of generating steering (Bell nonlocality) capability i.e., $\alpha_{\rm{steer}}=1$ ($\alpha_{\rm{Bell}}=1$) and $\beta_{\rm{steer}}=0.4641$ ($\beta_{\rm{Bell}}=0.1716$). 
Furthermore, since the target process is the process that experiences the ideal CPHASE gate with $\lambda=\pi$, 
the fidelity criterion obtained by the incapable (unable) process is equal to $F_{\mathcal{I}}\sim0.6830$ ($F_{\mathcal{U}}\sim0.8536$).
Thus, as shown in Figs.~\ref{fig_sb_a} to \ref{fig_sb_f}, 
the steering (Bell nonlocality) capability results obtained from the \textit{ibmq\_qasm\_simulator} and \textit{Amazon Braket local simulator} are similar to those of the ideal CPHASE gate process. For the real devices, however, the \textit{ibmq\_santiago} device has the capability to generate steering and Bell nonlocality only when the CPHASE shift is equal to $\pi$ since, in accordance with the fidelity criterion, they cannot be mimicked by an incapable process and unable process, respectively. The \textit{Amazon Braket Rigetti Aspen-9} (\textit{Aspen-M-1}) device is similarly identified to be capable of generating steering for CPHASE shifts of $\pi$ ($3/4\pi$, $\pi$, $5/4\pi$, and $3/2\pi$), as shown in Fig.~\ref{fig_sb_f}(a). 
The quantification and identification results were further investigated using a noise model created from the properties of the \textit{ibmq\_santiago} device. 
(The details of the noise modeling approach are presented in Appendix~\ref{app:IBMQNM}.) 
The simulation results are similar to the experimental results obtained on the \textit{ibmq\_santiago} device for all three capability indicators. 
Further noise models were constructed based on the properties of the \textit{Amazon Braket Rigetti Aspen-9} and \textit{Aspen-M-1} devices, respectively. (The noise models are described in Appendix~\ref{app:QCNMA}.) In this case, the simulated results deviated from the experimental results obtained on the corresponding real-world \textit{Amazon Braket Rigetti Aspen-9} and \textit{Aspen-M-1} devices (see Appendix~\ref{app:QCNMA} for the noise model of Amazon Braket).

\section{\label{sec:cao}Conclusion and Outlook}

In this work, we have investigated the problem of identifying and quantifying the quantum correlation generating capability of experimental processes. We have considered two particular quantum correlation generating processes, namely EPR steering generating process and Bell nonlocality generating process. We have defined both types of processes using a similar concept of quantum process capability theory~\cite{hsieh2017quantifying,kuo2019quantum}. We have proposed two measures and two identifiers (namely composition and robustness) for quantifying and identifying the capability of a process to generate steering or Bell nonlocality through the use of tomography and numerical methods. We have also presented two fidelity criteria for identifying faithful processes having the capability of generating steering and Bell nonlocality, respectively. The methods proposed herein are all based on classical mimicry methods with the concepts of local realism and classical dynamics.

Furthermore, we have presented and discussed the results obtained when using these approaches to quantify and identify the quantum correlation generation capability of a CPHASE gate process implemented on several real-world superconducting quantum computers, namely \textit{ibmq\_santiago}, \textit{Amazon Braket Rigetti Aspen-9} and \textit{Aspen-M-1}, and their corresponding simulators (with and without noise).

The experimentally feasible methods presented in this study for quantifying and identifying the steering and Bell nonlocality generating capabilities of a process provide a useful contribution toward the development of future cross-platform benchmarks for QIP tasks. In future work, we expect to extend the proposed formalism to characterize the multipartite correlation generation process further. Our methods may provide a way of quantifying quantum correlation generating processes and identifying the generation of multipartite nonclassical correlations for distributed QIP in entanglement-based quantum networks.

\acknowledgements

We thank Y.-N. Chen and H.-B. Chen for helpful comments and discussions.
We also appreciate S. Mangini's providing the calibration data of \textit{Amazon Braket Rigetti Aspen-9} downloaded on $30^{\text{th}}$ October 2021 to us.
This work was partially supported by the National Science and Technology Council, Taiwan, under Grant Numbers MOST 107-2628-M-006-001-MY4, MOST 111-2119-M-007-007, and MOST 111-2112-M-006-033.

\appendix

\section{\label{app:qptpm}TWO-QUBIT QUANTUM PROCESS TOMOGRAPHY AND PROCESS MATRIX}

A quantum system after process can be described by the process matrix $\chi_{\rm{expt}}$ with Eq.~(\ref{eq:1}). $\chi_{\rm{expt}}$ can then be used to reconstruct the process through operator-sum representation. For a $2$-qubit system with input states
$\rho_{{\rm{in}}}=\ket{\boldsymbol\phi_{i_mj_n}}\!\!\bra{\boldsymbol\phi_{i_mj_n}}$, where $m, n = \pm1$ when $i, j = 3$ and $m, n = 1$ when $i, j = 1, 2$ (defined in Sec.~\ref{sec:cpm}), the output states can be given explicitly as
\begin{equation}
\label{eq:3}
\rho_{{\rm{out}}}\equiv\chi_{\rm{expt}}(\rho_{{\rm{in}}})=\sum^{16}_{q=1}\sum^{16}_{r=1}\chi_{qr}E_q\rho_{{\rm{in}}}E^{\dagger}_r,
\end{equation}
where
\begin{equation}
\label{eq:4}
E_q=\bigotimes^2_{k=1}\ket{q_k}\!\!\bra{q_{k+2}},
\end{equation}
with $q=1+\sum^{4}_{i=1}q_i2^{i-1}$ for $q_i\in\{0,1\}$. In addition, $\ket{0}$ and $\ket{1}$ are defined in Sec.~\ref{sec:cpm}. To determine the coefficients $\chi_{qr}$ which constitute the process matrix $\chi_{\rm{expt}}$, we consider the following $16$ inputs:
\begin{equation}
\label{eq:5}
\rho_{\rm{in},q'}=E_{q'}=\bigotimes^2_{k=1}\ket{q'_k}\!\!\bra{q'_{k+2}},
\end{equation}
for $q'=1,2,...,16$. From Eq.~(\ref{eq:3}), the corresponding outputs are obtained as
\begin{equation}
\label{eq:6}
\chi_{\rm{expt}}(\rho_{\rm{in},q'})=\sum^1_{q_1=0}...\sum^1_{r_2=0}\bigotimes^2_{k=1}\ket{q_k}\!\!\bra{r_k}\chi_{g({\textbf{q}},q')h({\textbf{r}},q')},
\end{equation}
where $q'=1+\sum^{4}_{i=1}q'_i2^{i-1}$ for $q'_i\in\{0,1\},$ $\textbf{q}=(q_1,q_2),$ $\textbf{r}=(r_1,r_2),$
\begin{equation}
\label{eq:7}
g({\textbf{q}},q')=1+\sum^2_{i=1}q_i2^{i-1}+\sum^2_{i=1}q'_i2^{2+i-1},
\end{equation}
and
\begin{equation}
\label{eq:8}
h({\textbf{r}},q')=1+\sum^2_{i=1}r_i2^{i-1}+\sum^{4}_{i=2+1}q'_i2^{i-1}.
\end{equation}
Since the output $\chi(\rho_{\rm{in},q'})$ is determined using QST in Eq.~(\ref{eq:2}), we have full knowledge of the output matrix, i.e.,
\begin{equation}
\label{eq:9}
\rho'_{\rm{out},q'}=\chi_{\rm{expt}}(\rho_{\rm{in},q'})=\sum^1_{q_1=0}...\sum^1_{r_2=0}\bigotimes^2_{m=1}\ket{q_m}\!\!\bra{r_m}\rho^{(q')}_{\textbf{qr}}.
\end{equation}
Thus, all $16$ matrix elements $\rho^{(q')}_{\textbf{qr}}$ are determined. By comparing Eq.~(\ref{eq:6}) with Eq.~(\ref{eq:9}), the process matrix $\chi$ with $4\times4$ matrix elements can be obtained as
\begin{equation}
\label{eq:chi}
\chi_{g({\textbf{q}},q')h({\textbf{r}},q')}=\rho^{(q')}_{\textbf{qr}}.
\end{equation}
\hspace{1cm}From Eq.~(\ref{eq:4}), the $16$ operators, $E_q$, have the form
\begin{equation}
\label{eq:eqf}
E_q=\ket{q_1q_2}\bra{q_3q_4}
\end{equation}
where $E_1=\ket{00}\bra{00}$, $E_2=\ket{10}\bra{00}$, $E_3=\ket{01}\bra{00}$,..., and $E_{16}=\ket{11}\bra{11}$. To use experimentally preparable input states to obtain the process matrix, $E_q$, $q=1,2,...,16$, can be decomposed as a linear combination of the following density matrices of state: $\ket{\boldsymbol\phi_{i_mj_n}}\!\!\bra{\boldsymbol\phi_{i_mj_n}}$. The output matrix of $E_{k'}$, denoted as $\rho_{out,k'}$, can thus be represented by using the output density matrices of these states, denoted as $\rho'_{out,i_mj_n}=\chi_{\rm{expt}}(\ket{\boldsymbol\phi_{i_mj_n}}\!\!\bra{\boldsymbol\phi_{i_mj_n}})$ (It is worth noting that $\rho_{out,k'}$ is in $\{\ket{0}, \ket{1}\}$ basis, and $\rho'_{out,i_mj_n}$ is in the basis of Pauli matrices). That is, the process matrix can be written in the form
\begin{equation}
\label{eq:10}
\chi_{\rm{expt}} =\frac{1}{4} \left( \begin{matrix}
\rho_{\rm{out},1} & \rho_{\rm{out},5} & \rho_{\rm{out},9} & \rho_{\rm{out},13}
\\ \rho_{\rm{out},2} & \rho_{\rm{out},6} & \rho_{\rm{out},10} & \rho_{\rm{out},14}
\\ \rho_{\rm{out},3} & \rho_{\rm{out},7} & \rho_{\rm{out},11}& \rho_{\rm{out},15}
\\ \rho_{\rm{out},4} & \rho_{\rm{out},8} & \rho_{\rm{out},12}& \rho_{\rm{out},16}
\end{matrix} \right),
\end{equation}
where the constant $1/4$ is a normalization factor, and the diagonal elements are given as
\begin{equation}
\label{eq:outout}
\begin{aligned}
&\rho_{\rm{out},1}=\rho'_{\rm{out},3_13_1},\\
&\rho_{\rm{out},6}=\rho'_{\rm{out},3_{\text-1}3_1},\\
&\rho_{\rm{out},11}=\rho'_{\rm{out},3_13_{\text-1}},\\
&\rho_{\rm{out},16}=\rho'_{\rm{out},3_{\text-1}3_{\text-1}}.\\
\end{aligned}
\end{equation}
The other elements have the forms
\begin{equation}
\begin{aligned}
\rho_{\rm{out},2}= &\rho'_{\rm{out},1_13_1}\!\!-\!i\rho'_{\rm{out},2_13_1}\!\!-\!\frac{e^{-i\pi/4}}{\sqrt{2}}(\rho'_{\rm{out},3_13_1}\!+\!\rho'_{\rm{out},3_{\text-1}3_1}\!),\\
\rho_{\rm{out},3}= &\rho'_{\rm{out},3_11_1}\!\!-\!i\rho'_{\rm{out},3_12_1}\!\!-\!\frac{e^{-i\pi/4}}{\sqrt{2}}(\rho'_{\rm{out},3_13_1}\!+\!\rho'_{\rm{out},3_13_{\text-1}}\!),\\
\rho_{\rm{out},4}= &\rho'_{\rm{out},1_11_1}\!-\!i\rho'_{\rm{out},1_12_1}\!-\!\frac{e^{-i\pi/4}}{\sqrt{2}}(\rho'_{\rm{out},1_13_1}\!\!+\!\!\rho'_{\rm{out},1_13_{\text-1}})\\
&-\!i[\rho'_{\rm{out},2_11_1}\!\!\!-\!i\rho'_{\rm{out},2_12_1}\!\!\!-\!\frac{e^{-i\pi/4}}{\sqrt{2}}\!(\rho'_{\rm{out},2_13_1}\!\!\!+\!\!\rho'_{\rm{out},2_13_{\text-1}}\!)\!]\\
&-\!\frac{e^{-i\pi/4}}{\sqrt{2}}\{\![\rho'_{\rm{out},3_11_1}\!\!\!+\!\!\rho'_{\rm{out},3_{\text-1}1_1}\!\!\!-\!i(\rho'_{\rm{out},3_12_1}\!\!\!+\!\!\rho'_{\rm{out},3_{\text-1}2_1}\!)\!]\\
&-\!\frac{e^{-i\pi/4}}{\sqrt{2}}(\rho'_{\rm{out},3_13_1}\!\!+\!\!\rho'_{\rm{out},3_13_{\text-1}}\!\!+\!\!\rho'_{\rm{out},3_{\text-1}3_1}\!\!+\!\!\rho'_{\rm{out},3_{\text-1}3_{\text-1}}\!)\!\},\\
\rho_{\rm{out},5}= &\rho_{\rm{out},2}^\dag,\\
\nonumber
\end{aligned}
\end{equation}
\begin{equation}
\begin{aligned}
\rho_{\rm{out},7}= &\rho'_{\rm{out},1_11_1}\!-\!i\rho'_{\rm{out},1_12_1}\!-\!\frac{e^{-i\pi/4}}{\sqrt{2}}(\rho'_{\rm{out},1_13_1}\!\!+\!\!\rho'_{\rm{out},1_13_{\text-1}})\\
&+\!i[\rho'_{\rm{out},2_11_1}\!\!\!-\!i\rho'_{\rm{out},2_12_1}\!\!\!-\!\!\frac{e^{-i\pi/4}}{\sqrt{2}}\!(\rho'_{\rm{out},2_13_1}\!\!\!+\!\!\rho'_{\rm{out},2_13_{\text-1}}\!)\!]\\
&-\!\frac{e^{i\pi/4}}{\sqrt{2}}\{\![\rho'_{\rm{out},3_11_1}\!\!\!+\!\!\rho'_{\rm{out},3_{\text-1}1_1}\!\!\!-\!i(\rho'_{\rm{out},3_12_1}\!\!\!+\!\!\rho'_{\rm{out},3_{\text-1}2_1}\!)\!]\\
&-\!\frac{e^{-i\pi/4}}{\sqrt{2}}(\rho'_{\rm{out},3_13_1}\!\!+\!\!\rho'_{\rm{out},3_13_{\text-1}}\!\!+\!\!\rho'_{\rm{out},3_{\text-1}3_1}\!\!+\!\!\rho'_{\rm{out},3_{\text-1}3_{\text-1}}\!)\!\},\\
\rho_{\rm{out},8}= &\rho'_{\rm{out},3_{\text-1}1_1}\!\!-i\rho'_{\rm{out},3_{\text-1}2_1}\!\!-\frac{e^{-i\pi/4}}{\sqrt{2}}(\rho'_{\rm{out},3_{\text-1}3_1}\!\!+\!\rho'_{\rm{out},3_{\text-1}3_{\text-1}}\!),\\
\rho_{\rm{out},9}= &\rho_{\rm{out},3}^\dag,\\
\rho_{\rm{out},10}= &\rho_{\rm{out},7}^\dag,\\
\rho_{\rm{out},12}= &\rho'_{\rm{out},1_13_{\text-1}}\!\!-i\rho'_{\rm{out},2_13_{\text-1}}\!\!-\frac{e^{-i\pi/4}}{\sqrt{2}}(\rho'_{\rm{out},3_13_{\text-1}}\!\!+\!\rho'_{\rm{out},3_{\text-1}3_{\text-1}}\!),\\
\rho_{\rm{out},13}= &\rho_{\rm{out},4}^\dag,\\
\rho_{\rm{out},14}= &\rho_{\rm{out},8}^\dag,\\
\rho_{\rm{out},15}= &\rho_{\rm{out},12}^\dag.\\
\label{eq:outout2} %\nonumber
\end{aligned}
\end{equation}

\section{\label{app:IBMQNM}NOISE MODELING OF IBM Q}

IBM~Q~\cite{ibmq} provides several simulators for use, including the \textit{ibmq\_qasm\_simulator}, \textit{simulator\_statevector}, \textit{simulator\_extended\_stabilizer}, and \textit{simulator\_mps}, where these simulators can simulate circuits of up to $32$, $32$, $63$, and $100$ qubits, respectively. However only the \textit{ibmq\_qasm\_simulator} and \textit{simulator\_statevector} devices provide noise modeling~\cite{sourceCodeCH}. Thus, in this study, we purposely chose the \textit{ibmq\_qasm\_simulator} device since it not only supports noise modeling, but is also available for use without connecting to IBM~Q's service.
Noise modeling is the process of building a noise model to simulate a quantum circuit in the presence of errors. It is possible to construct custom noise models for simulators or automatically generate a basic noise model from an IBM Q device. Moreover, a simplified approximate noise model can be generated automatically from the properties of a real device. 
The noise model was constructed using the properties obtained from the \textit{ibmq\_santiago} device on $8^{\text{th}}$ October $2021$.

\section{\label{app:QCNMA}QUANTIFICATION AND IDENTIFICATION OF CONTROLLED-PHASE GATE ON NOISE MODEL OF AMAZON BRAKET} 

In Sec.~\ref{sec:QCSQC}, we used the noise model in the IBM~Q backend to simulate the real-world quantum computer. 
In this Appendix, we simulate the Amazon Braket quantum devices and evaluate their performance by replacing the parameters in the noise model of the IBM~Q backend with the calibration data of the Amazon Braket backend. 
In particular, we show how the noisy \textit{ibmq\_qasm\_simulator} device is combined with the calibration data of the \textit{Amazon Braket Rigetti Aspen-9} and \textit{Aspen-M-1} devices, respectively, to obtain the results shown in Figs.~\ref{fig_sb_a}, \ref{fig_sb_b}, and \ref{fig_sb_f}.

For the \textit{ibmq\_santiago} noise model, the most important parameters include $T_{\rm{1}}$, $T_{\rm{2}}$, the \text{readout error}, \textit{prob\_meas0\_prep1}, \textit{prob\_meas1\_prep0}, the \text{gate length}, and the \text{gate error}. (Note that the values of these parameters can be taken directly from the IBM Q program.)
From the source code~\cite{sourceCodeCH} of the noise model of the IBM Q backend, we found that the noise model is generated based on the 1-qubit and 2-qubit gate errors (which consist of a depolarizing error followed by a thermal relaxation error and describe a CPTP $N$-qubit gate) and the single-qubit readout error on all the measurements. The error parameters of the noise model are tuned based on the parameters $T_{\rm{1}}$, $T_{\rm{2}}$, \text{frequency}, \text{readout error}, \textit{prob\_meas0\_prep1}, \textit{prob\_meas1\_prep0}, \text{gate length}, and \text{gate error}. Therefore, it is necessary to collect data for all these parameters to construct the noise models for the Amazon Braket Rigetti devices.

For the Amazon Braket Rigetti quantum computer, the calibration data provided by Amazon Braket Rigetti are $T_{\rm{1}}$, $T_{\rm{2}}$, the readout fidelity, the gate time, the \textit{RB\_Fidelity}, and the 2-qubit gate fidelity. 
We note that all these data can be obtained directly from the Amazon Braket console~\cite{awsCH} and Rigetti~\cite{rigettiCH} websites.
We note also that there are three common parameters in the noise models of the IBM Q and Amazon Braket devices, respectively.
For example, $T_{\rm{1}}$ is the energy relaxation time, i.e., the time scale of the decay of a qubit from the excited state to the ground state, which is related to the amplitude damping noise. The effect of amplitude damping in the Bloch representation as the Bloch vector transformation~\cite{nielsen2002quantum} is given as \begin{equation}
(r_x, r_y, r_z) \mapsto (r_x \sqrt{1-\gamma}, r_y \sqrt{1-\gamma}, r_z (1-\gamma) + \gamma),
\end{equation} 
where 
\begin{equation}
\gamma = 1 - \exp{-\frac{t}{T_1}},
\end{equation} and \textit{t} is time.
In addition, $T_{\rm{2}}$ is the dephasing time, i.e., the time scale required for the decoherence of a qubit from the coherent state to a completely mixed state, which is related to the phase damping noise. The effect of phase damping in the Bloch representation as the Bloch vector transformation~\cite{nielsen2002quantum} is given as \begin{equation}
(r_x, r_y, r_z) \mapsto (r_x \sqrt{1-\lambda}, r_y \sqrt{1-\lambda}, r_z),
\end{equation} 
where 
\begin{equation}
\lambda = 1 - \exp{-\frac{t}{(T_2/2)}},
\end{equation} and \textit{t} is time.
If the amplitude damping noise and phase damping noise are both applied, and the latter is followed by the former, then 
\begin{equation}\label{eq:C5}
T_2 \leq 2 \times T_1.
\end{equation}
An inspection of the source code~\cite{sourceCodeCH} of the noise model of the IBM Q backend showed the existence of several constraints which ensure that Eq.~(\ref{eq:C5}) holds.
The \text{gate length} of IBM Q is equivalent to gate time, and represents the duration for which the gate operates on one or two specific qubits.
For all three common parameters, we simply replaced the parameter values of IBM Q with those of Amazon Braket in the same units. 

For the parameters which are not common to both models, e.g., the readout fidelity of a single qubit, the \textit{RB\_Fidelity} and the 2-qubit gate fidelity, we derived specific relations to relate the calibration data of the Amazon Braket device to the parameters of the IBM Q.
%\textcolor{blue}{The first one is the readout fidelity of a single qubit.} 
In general, a readout error causes the measurement result for a qubit in the $\ket{0}$ state to be given as the measurement result for a qubit in the $\ket{1}$ state, and vice versa. 
The \textit{prob\_meas0\_prep1} parameter of IBM Q gives the probability that the measurement result of $\ket{1}$ is that of $\ket{0}$, while the \textit{prob\_meas1\_prep0} parameter gives the probability that the measurement result of $\ket{0}$ is that of $\ket{1}$.
In the source code~\cite{sourceCodeCH} of the noise model of the IBM Q backend, \textit{prob\_meas0\_prep1} and \textit{prob\_meas1\_prep0} are adopted preferentially rather than the readout error (if provided).
Based on our reading of the Rigetti pyQuil website~\cite{rigetti2CH}, we surmised that the classical readout bit-flip error was roughly symmetric for the simulation.
In other words, the readout errors \textit{prob\_meas0\_prep1} and \textit{prob\_meas1\_prep0} of the noise model were the same. Thus, we set the readout error equal to $1$ minus the readout fidelity as a reasonable approximation.
The \textit{RB\_Fidelity} is the single-qubit randomized benchmarking fidelity~\cite{knill2008randomized} of the individual gate operation.
The 2-qubit gate fidelity included the C-Phase gate (denoted in Amazon Braket console~\cite{awsCH}), i.e., CPHASE gate~[Eq.~(\ref{CPHASEEEE})] fidelity, the XY gate~\cite{Abrams2019} fidelity, and the CZ gate (i.e., controlled-Z gate) fidelity.
The state fidelity corresponding to the number of computational gates was given by 1 minus the probability of error in Fig.~1 of Ref.~\cite{knill2008randomized}.
Thus, for the noise models of the \textit{Amazon Braket Rigetti Aspen-9} and \textit{Aspen-M-1} devices, we assigned 1 minus \textit{RB\_Fidelity} to the value of the 1-qubit gate error and 1 minus the C-Phase gate fidelity to the value of the 2-qubit gate error.

\begin{table*}%[b]%The best place to locate the table environment is directly after its first reference in text
\caption{\label{tab:table1}Comparison among parameters of different noise models. First, we used the IBM~Q \textit{ibmq\_santiago} quantum computer with qubit 3 and qubit 4 in October 2021 and archived its calibration parameters on $8^{\text{th}}$ October 2021. Second, we used the \textit{Amazon Braket Rigetti Aspen-9} quantum computer with qubit 10 and qubit 17 in July 2021 and archived its calibration parameters on $30^{\text{th}}$ October and $18^{\text{th}}$ November 2021. Finally, we used the \textit{Amazon Braket Rigetti Aspen-M-1} quantum computer with qubit 15 and qubit 16 on 
$23^{\text{rd}}$ and $30^{\text{th}}$ of March 2022, respectively, and archived the calibration parameters on the same dates.}
\begin{ruledtabular}
\begin{tabular}{lccc}
&IBM~Q \textit{ibmq\_santiago}& \textit{Amazon Braket Aspen-9} & \textit{Amazon Braket Aspen-M-1}\\ 
& qubit 3 \& 4 & qubit 10 \& 17 & qubit 15 \& 16 \\ 
\hline
$T_{\rm{1}}$ ($\mu\text{s}$) & 106.2285 \& 44.8018 & 26.43 \& 28.88\footnotemark[1]  & 50.79 \& 40.518 \\
$T_{\rm{2}}$ ($\mu\text{s}$) & 82.9952 \& 88.0221 & 21.62 \& 24.14\footnotemark[1] & 60.606 \& 65.261 \\
\textit{prob\_meas0\_prep1}, \textit{prob\_meas1\_prep0} & 0.0082, 0.0044 \& 0.0346, 0.0112  & & \\
readout error & 0.0063 \& 0.0229 & 1 - 0.957 \& 1 - 0.939\footnotemark[1] & 1 - 0.983 \& 1 - 0.987 \\
% &&& 1 - 0.981 \& 1 - 0.974 \\
1-qubit, 2-qubit gate time ($\text{ns}$) & 35.5556, 376.8889 & 48, 168 & 40, 180 \\
1-qubit gate error & 0.0002 \& 0.0003 & 1 - 0.9989 \& 1 - 0.9993\footnotemark[1] & 1 - 0.9987 \& 1 - 0.99947 \\
2-qubit gate error & 0.0056 & 1 - 0.97955\footnotemark[2] & 1 - 0.98996 \\
% &&& 1 - 0.98250 \\
\end{tabular}
\end{ruledtabular}
\footnotetext[1]{
%Stefano Mangini~\cite{mangini2021qubit} kindly provided the calibration data of \textit{Amazon Braket Rigetti Aspen-9} downloaded on 31 October 2021 from the quantum cloud services website of Rigetti Computing~\cite{rigetti3CH}.
The calibration data of \textit{Amazon Braket Rigetti Aspen-9}~\cite{mangini2021qubit} is downloaded from the quantum cloud services website of Rigetti Computing~\cite{rigetti3CH} on $30^{\text{th}}$ October 2021.
}
\footnotetext[2]{Due to the lack of C-Phase gate fidelity in the calibration data of \textit{Amazon Braket Rigetti Aspen-9}~\cite{mangini2021qubit} downloaded on $30^{\text{th}}$ October 2021, we used the data from the website of Amazon Braket console~\cite{awsCH} on $18^{\text{th}}$ November 2021 instead.}
\end{table*}

Table~\ref{tab:table1} shows the calibration data used to construct the noise models for the \textit{Amazon Braket Rigetti Aspen-9} and \textit{Aspen-M-1} devices. (Note that the original parameters of the IBM Q \textit{ibmq\_santiago} device are also shown for reference purposes.)
The 1-qubit gates of IBM Q \textit{ibmq\_santiago} include the Identity gate, the sx ($\sqrt{x}$) gate, and the Pauli-X gate, while the 2-qubit gate is the CNOT gate (cx3\_4) operated on qubit 3 and qubit 4. The 2-qubit gate fidelities of \textit{Amazon Braket Rigetti Aspen-M-1} and \textit{Aspen-9} are the C-Phase gate fidelities, and the gate times are derived directly from the websites of Amazon Braket console~\cite{awsCH} and Rigetti~\cite{rigettiCH}, respectively. The data for \textit{Amazon Braket Rigetti Aspen-M-1} were collected on $23^{\text{rd}}$ March $2022$ for the Bell nonlocality generating test. 
For the steering generating test, the data were collected on $30^{\text{th}}$ March $2022$. The readout fidelity was transformed to $0.981$ for qubit $15$ and $0.974$ for qubit $16$, while the \text{2-qubit gate fidelity} was transformed to $0.9825$. Since the calibration data of the Amazon Braket devices changed over time, the data were updated in such a way that the noise model of the Amazon Braket device exhibited the same time-varying behavior as the original noise model of IBM Q.


\begin{thebibliography}{100}

\bibitem{feynman2018simulating}R. P. Feynman, Simulating physics with computers, Int. J. Theor. Phys. \textbf{21}, 467 (1982).

\bibitem{deutsch1985quantum}D. Deutsch, Quantum theory, the Church--Turing principle and the universal quantum computer, Proc. R. Soc. Lond. A \textbf{400}, 97 (1985).

\bibitem{lloyd1996universal}S. Lloyd, Universal quantum simulators, Science \textbf{273}, 1073 (1996).

\bibitem{divincenzo2000physical}D. P. DiVincenzo, The physical implementation of quantum computation, Fortschr. Phys. \textbf{48}, 771 (2000).

\bibitem{nielsen2002quantum}M. A. Nielsen and I. L. Chuang, \textit{Quantum Computation and Quantum Information} (Cambridge University Press, Cambridge, UK, 2000).

\bibitem{georgescu2014quantum}I. M. Georgescu, S. Ashhab, and F. Nori, Quantum simulation, Rev. Mod. Phys. \textbf{86}, 153 (2014).

\bibitem{preskill2018quantum}J. Preskill, Quantum computing in the NISQ era and beyond, Quantum \textbf{2}, 79 (2018).

\bibitem{bruzewicz2019trapped}C. D. Bruzewicz, J. Chiaverini, R. McConnell, and J. M. Sage, Trapped-ion quantum computing: progress and challenges, Appl. Phys. Rev. \textbf{6}, 021314 (2019).

\bibitem{kjaergaard2020superconducting}M. Kjaergaard, M. E. Schwartz, J. Braum{\"u}ller, P. Krantz, J. I.-J. Wang, S. Gustavsson, and W. D. Oliver, Superconducting qubits: current state of play, Annu. Rev. Condens. Matter Phys. \textbf{11}, 369 (2020).

\bibitem{kane1998silicon}B. E. Kane, A silicon-based nuclear spin quantum computer, Nature \textbf{393}, 133 (1998). %18

\bibitem{ladd2002all}T. D. Ladd, J. R. Goldman, F. Yamaguchi, Y. Yamamoto, E. Abe, and K. M. Itoh, All-silicon quantum computer, Phys. Rev. Lett. \textbf{89}, 017901 (2002). %19

\bibitem{zwanenburg2013silicon}F. A. Zwanenburg, A. S. Dzurak, A. Morello, M. Y. Simmons, L. C. L. Hollenberg, G. Klimeck, S. Rogge, S. N. Coppersmith, and M. A. Eriksson, Silicon quantum electronics, Rev. Mod. Phys. \textbf{85}, 961 (2013).

\bibitem{xiang2013hybrid}Z.-L. Xiang, S. Ashhab, J. Q. You, and F. Nori, Hybrid quantum circuits: superconducting circuits interacting with other quantum systems, Rev. Mod. Phys. \textbf{85}, 623 (2013). %16

\bibitem{madsen2022quantum}L. S. Madsen \textit{et al.}, Quantum computational advantage with a programmable photonic processor, Nature \textbf{606}, 75 (2022).

\bibitem{nakamura1999coherent}Y. Nakamura, Y. A. Pashkin, and J. S. Tsai, Coherent control of macroscopic quantum states in a single-Cooper-pair box, Nature \textbf{398}, 786 (1999). %15

\bibitem{makhlin2001quantum}Y. Makhlin, G. Sch{\"o}n, and A. Shnirman, Quantum-state engineering with Josephson-junction devices, Rev. Mod. Phys. \textbf{73}, 357 (2001). %10

\bibitem{krantz2019quantum}P. Krantz, M. Kjaergaard, F. Yan, T. P. Orlando, S. Gustavsson, and W. D. Oliver, A quantum engineer's guide to superconducting qubits, Appl. Phys. Rev. \textbf{6}, 021318 (2019). %17

\bibitem{cirac1995quantum}J. I. Cirac and P. Zoller, Quantum computations with cold trapped ions, Phys. Rev. Lett. \textbf{74}, 4091 (1995). % 13

\bibitem{haffner2008quantum}H. H{\"a}ffner, C.F. Roos, and R. Blatt, Quantum computing with trapped ions, Phys. Rep. \textbf{469}, 155 (2008).

\bibitem{ibmq}IBM Q Experience, \url{https://quantum-computing.ibm.com}.

\bibitem{aws}Amazon Braket, \url{https://aws.amazon.com/braket/}.

\bibitem{knill2008randomized}E. Knill, D. Leibfried, R. Reichle, J. Britton, R. B. Blakestad, J. D. Jost, C. Langer, R. Ozeri, S. Seidelin, and D. J. Wineland, Randomized benchmarking of quantum gates, Phys. Rev. A \textbf{77}, 012307 (2008).

\bibitem{blume2013robust}R. Blume-Kohout, J. K. Gamble, E. Nielsen, J. Mizrahi, J. D. Sterk, and P. Maunz, Robust, self-consistent, closed-form tomography of quantum logic gates on a trapped ion qubit, arXiv:1310.4492. % 22  %(October 17, 2013)

\bibitem{chen2014qubit}Y. Chen et al., Qubit architecture with high coherence and fast tunable coupling, Phys. Rev. Lett. \textbf{113}, 220502 (2014). % 24

\bibitem{veldhorst2014addressable}M. Veldhorst et al., An addressable quantum dot qubit with fault-tolerant control-fidelity, Nat. Nanotechnol. \textbf{9}, 981 (2014). % 28

\bibitem{blume2017demonstration}R. Blume-Kohout, J. K. Gamble, E. Nielsen, K. Rudinger, J. Mizrahi, K. Fortier, and P. Maunz, Demonstration of qubit operations below a rigorous fault tolerance threshold with gate set tomography, Nat. Commun. \textbf{8}, 14485 (2017).   % 23

\bibitem{benedetti2019generative}M. Benedetti, D. Garcia-Pintos, O. Perdomo, V. Leyton-Ortega, Y. Nam, and A. Perdomo-Ortiz, A generative modeling approach for benchmarking and training shallow quantum circuits, npj Quantum Inf. \textbf{5}, 45 (2019).  % 21

\bibitem{cross2019validating}A. W. Cross, L. S. Bishop, S. Sheldon, P. D. Nation, and J. M. Gambetta, Validating quantum computers using randomized model circuits, Phys. Rev. A \textbf{100}, 032328 (2019). % 25

\bibitem{huang2019fidelity}W. Huang et al., Fidelity benchmarks for two-qubit gates in silicon, Nature \textbf{569}, 532 (2019). % 26

\bibitem{xue2019benchmarking}X. Xue, T. F. Watson, J. Helsen, D. R. Ward, D. E. Savage, M. G. Lagally, S. N. Coppersmith, M. A. Eriksson, S. Wehner, and L. M. K. Vandersypen, Benchmarking gate fidelities in a Si/SiGe two-qubit device, Phys. Rev. X \textbf{9}, 021011 (2019). % 29

\bibitem{jurcevic2021demonstration}P. Jurcevic et al., Demonstration of quantum volume 64 on a superconducting quantum computing system, Quantum Sci. Technol. \textbf{6}, 025020 (2021). % 27

\bibitem{knill2001benchmarking}E. Knill, R. Laflamme, R. Martinez, and C. Negrevergne, Benchmarking quantum computers: the five-qubit error correcting code, Phys. Rev. Lett. \textbf{86}, 5811 (2001). % 31

\bibitem{mccaskey2019quantum}A. J. McCaskey, Z. P. Parks, J. Jakowski, S. V. Moore, T. D. Morris, T. S. Humble, and R. C. Pooser, Quantum chemistry as a benchmark for near-term quantum computers, npj Quantum Inf. \textbf{5}, 99 (2019). % 32

\bibitem{wright2019benchmarking}K. Wright et al., Benchmarking an 11-qubit quantum computer, Nat. Commun. \textbf{10}, 5464 (2019). % 33

\bibitem{harrigan2021quantum}M. P. Harrigan et al., Quantum approximate optimization of non-planar graph problems on a planar superconducting processor, Nat. Phys. \textbf{17}, 332 (2021). % 30

\bibitem{einstein1935can}A. Einstein, B. Podolsky, and N. Rosen, Can quantum-mechanical description of physical reality be considered complete?, Phys. Rev. \textbf{47}, 777 (1935).

\bibitem{horodecki2009quantum}R. Horodecki, P. Horodecki, M. Horodecki, and K. Horodecki, Quantum entanglement, Rev. Mod. Phys. \textbf{81}, 865 (2009).

\bibitem{schrodinger1935discussion}E. Schr{\"o}dinger, Discussion of probability relations between separated systems, in {\em Mathematical Proceedings of the Cambridge Philosophical Society, 1935}, p. 555.
% Math. Proc. Camb. Phil. Soc. \textbf{31}, 555 (1935).

\bibitem{wiseman2007steering}H. M. Wiseman, S. J. Jones, and A. C. Doherty, Steering, entanglement, nonlocality, and the Einstein-Podolsky-Rosen paradox, Phys. Rev. Lett. \textbf{98}, 140402 (2007).

\bibitem{uola2019quantum}R. Uola, A. C. S. Costa, H. C. Nguyen, and O. G{\"u}hne, Quantum steering, Rev. Mod. Phys. \textbf{92}, 015001 (2020).

\bibitem{bell1964einstein}J. S. Bell, On the Einstein Podolsky Rosen paradox, Physics \textbf{1}, 195 (1964).
% Physics Physique Fizika
%Physics  % https://journals.aps.org/pra/references/10.1103/PhysRevA.54.3824
%Physics  % https://journals.aps.org/pra/references/10.1103/PhysRevA.66.032110

\bibitem{clauser1969proposed}J. F. Clauser, M. A. Horne, A. Shimony, and R. A. Holt, Proposed experiment to test local hidden-variable theories, Phys. Rev. Lett. \textbf{23}, 880 (1969).

\bibitem{brunner2014bell}N. Brunner, D. Cavalcanti, S. Pironio, V. Scarani, and S. Wehner, Bell nonlocality, Rev. Mod. Phys. \textbf{86}, 419 (2014).

\bibitem{bennett1993teleporting}C. H. Bennett, G. Brassard, C. Cr{\'e}peau, R. Jozsa, A. Peres, and W. K. Wootters, Teleporting an unknown quantum state via dual classical and Einstein-Podolsky-Rosen channels, Phys. Rev. Lett. \textbf{70}, 1895 (1993).

\bibitem{bouwmeester1997experimental}D. Bouwmeester, J.-W. Pan, K. Mattle, M. Eibl, H. Weinfurter, and A. Zeilinger, Experimental quantum teleportation, Nature \textbf{390}, 575 (1997).

\bibitem{hillery1999quantum}M. Hillery, V. Bu{\v{z}}ek, and A. Berthiaume, Quantum secret sharing, Phys. Rev. A \textbf{59}, 1829 (1999).

\bibitem{chen2005experimental}Y.-A. Chen, A.-N. Zhang, Z. Zhao, X.-Q. Zhou, C.-Y. Lu, C.-Z. Peng, T. Yang, and J.-W. Pan, Experimental quantum secret sharing and third-man quantum cryptography, Phys. Rev. Lett. \textbf{95}, 200502 (2005).

\bibitem{PhysRevLett.86.5188}R. Raussendorf and H. J. Briegel, A one-way quantum computer, Phys. Rev. Lett. \textbf{86}, 5188 (2001).

\bibitem{raussendorf2003measurement}R. Raussendorf, D. E. Browne, and H. J. Briegel, Measurement-based quantum computation on cluster states, Phys. Rev. A \textbf{68}, 022312 (2003).

\bibitem{walther2005experimental}P. Walther, K. J. Resch, T. Rudolph, E. Schenck, H. Weinfurter, V. Vedral, M. Aspelmeyer, and A. Zeilinger, Experimental one-way quantum computing, Nature \textbf{434}, 169 (2005).

\bibitem{branciard2012one}C. Branciard, E. G. Cavalcanti, S. P. Walborn, V. Scarani, and H. M. Wiseman, One-sided device-independent quantum key distribution: security, feasibility, and the connection with steering, Phys. Rev. A \textbf{85}, 010301 (2012).

\bibitem{ekert1991quantum}A. K. Ekert, Quantum cryptography based on Bell's theorem, Phys. Rev. Lett. \textbf{67}, 661 (1991).

\bibitem{brukner2004bell}{\v{C}}. Brukner, M. {\.Z}ukowski, J.-W. Pan, and A. Zeilinger, Bell's inequalities and quantum communication complexity, Phys. Rev. Lett. \textbf{92}, 127901 (2004).

\bibitem{buhrman2010nonlocality}H. Buhrman, R. Cleve, S. Massar, and R. de Wolf, Nonlocality and communication complexity, Rev. Mod. Phys. \textbf{82}, 665 (2010).

\bibitem{de2014nonlocality}J. I de Vicente, On nonlocality as a resource theory and nonlocality measures, J. Phys. A: Math. Theor. \textbf{47}, 424017 (2014).

\bibitem{piani2008no}M. Piani, P. Horodecki, and R. Horodecki, No-local-broadcasting theorem for multipartite quantum correlations, Phys. Rev. Lett. \textbf{100}, 090502 (2008).

\bibitem{ma2016converting}J. Ma, B. Yadin, D. Girolami, V. Vedral, and M. Gu, Converting coherence to quantum correlations, Phys. Rev. Lett. \textbf{116}, 160407 (2016).

\bibitem{vogel1989determination}K. Vogel and H. Risken, Determination of quasiprobability distributions in terms of probability distributions for the rotated quadrature phase, Phys. Rev. A \textbf{40}, 2847 (1989). % 60

\bibitem{Horodecki_1996}M. Horodecki, P. Horodecki, and R. Horodecki, Separability of mixed states: necessary and sufficient conditions, Phys. Lett. A \textbf{223}, 1 (1996). % 56

\bibitem{peres1996separability}A. Peres, Separability criterion for density matrices, Phys. Rev. Lett. \textbf{77}, 1413 (1996). % 58

\bibitem{leonhardt1997measuring}U. Leonhardt, \textit{Measuring the quantum state of light} (Cambridge University Press, 1997). % 57

\bibitem{PhysRevLett.80.2245}W. K. Wootters, Entanglement of formation of an arbitrary state of two qubits, Phys. Rev. Lett. \textbf{80}, 2245 (1998). % 61

\bibitem{PhysRevA.65.032314}G. Vidal and R. F. Werner, Computable measure of entanglement, Phys. Rev. A \textbf{65}, 032314 (2002). % 59

\bibitem{cavalcanti2009experimental}E. G. Cavalcanti, S. J. Jones, H. M. Wiseman, and M. D. Reid, Experimental criteria for steering and the Einstein-Podolsky-Rosen paradox, Phys. Rev. A \textbf{80}, 032112 (2009).

\bibitem{skrzypczyk2014quantifying}P. Skrzypczyk, M. Navascu{\'e}s, and D. Cavalcanti, Quantifying Einstein-Podolsky-Rosen steering, Phys. Rev. Lett. \textbf{112}, 180404 (2014).

\bibitem{gallego2015resource}R. Gallego and L. Aolita, Resource theory of steering, Phys. Rev. X \textbf{5}, 041008 (2015).

\bibitem{piani2015necessary}M. Piani and J. Watrous, Necessary and sufficient quantum information characterization of Einstein-Podolsky-Rosen steering, Phys. Rev. Lett. \textbf{114}, 060404 (2015).

\bibitem{eberhard1993background}P. H. Eberhard, Background level and counter efficiencies required for a loophole-free Einstein-Podolsky-Rosen experiment, Phys. Rev. A \textbf{47}, R747 (1993). % 71

\bibitem{popescu1994quantum}S. Popescu and D. Rohrlich, Quantum nonlocality as an axiom, Found. Phys. \textbf{24}, 379 (1994). % 78

\bibitem{pironio2003violations}S. Pironio, Violations of Bell inequalities as lower bounds on the communication cost of nonlocal correlations, Phys. Rev. A \textbf{68}, 062102 (2003). % 77

\bibitem{toner2003communication}B. F. Toner and D. Bacon, Communication cost of simulating Bell correlations, Phys. Rev. Lett. \textbf{91}, 187904 (2003). % 80

\bibitem{acin2005optimal}A. Ac{\'\i}n, R. Gill, and N. Gisin, Optimal Bell tests do not require maximally entangled states, Phys. Rev. Lett. \textbf{95}, 210402 (2005). % 66

\bibitem{van2005statistical}W. van Dam, R. D. Gill, and P. D. Grunwald, The statistical strength of nonlocality proofs, IEEE Trans. Inf. Theory \textbf{51}, 2812 (2005). % 81

\bibitem{junge2010operator}M. Junge, C. Palazuelos, D. P{\'e}rez-Garc{\'\i}a, I. Villanueva, and M. M. Wolf, Operator space theory: a natural framework for Bell inequalities, Phys. Rev. Lett. \textbf{104}, 170405 (2010). % 75

\bibitem{hall2011relaxed}M. J. W. Hall, Relaxed Bell inequalities and Kochen-Specker theorems, Phys. Rev. A \textbf{84}, 022102 (2011). % 74

\bibitem{chaves2012multipartite}R. Chaves, D. Cavalcanti, L. Aolita, and A. Ac{\'\i}n, Multipartite quantum nonlocality under local decoherence, Phys. Rev. A \textbf{86}, 012108 (2012). % 69

\bibitem{chaves2015unifying}R. Chaves, R. Kueng, J. B. Brask, and D. Gross, Unifying framework for relaxations of the causal assumptions in Bell's theorem, Phys. Rev. Lett. \textbf{114}, 140403 (2015). % 70

\bibitem{fonseca2015measure}E. A. Fonseca and F. Parisio, Measure of nonlocality which is maximal for maximally entangled qutrits, Phys. Rev. A \textbf{92}, 030101(R) (2015). % 72

\bibitem{montina2016information}A. Montina and S. Wolf, Information-based measure of nonlocality, New J. Phys. \textbf{18}, 013035 (2016). % 76

\bibitem{ringbauer2016experimental}M. Ringbauer, C. Giarmatzi, R. Chaves, F. Costa, A. G. White, and A. Fedrizzi, Experimental test of nonlocal causality, Sci. Adv. \textbf{2}, e1600162 (2016). % 79

\bibitem{brask2017bell}J. B. Brask and R. Chaves, Bell scenarios with communication, J. Phys. A: Math. Theor. \textbf{50}, 094001 (2017). % 67

\bibitem{gallego2017nonlocality}R. Gallego and L. Aolita, Nonlocality free wirings and the distinguishability between Bell boxes, Phys. Rev. A \textbf{95}, 032118 (2017). % 73

\bibitem{brito2018quantifying}S. G. A. Brito, B. Amaral, and R. Chaves, Quantifying Bell nonlocality with the trace distance, Phys. Rev. A \textbf{97}, 022111 (2018). % 68

\bibitem{campbell2010optimal}E. T. Campbell, Optimal entangling capacity of dynamical processes, Phys. Rev. A \textbf{82}, 042314 (2010).

\bibitem{moravvcikova2010entanglement}L. Morav{\v{c}}{\'\i}kov{\'a} and M. Ziman, Entanglement-annihilating and entanglement-breaking channels, J. Phys. A: Math. Theor. \textbf{43}, 275306 (2010).

\bibitem{mao2020experimentally}Y. Mao, Y.-Z. Zhen, H. Liu, M. Zou, Q.-J. Tang, S.-J. Zhang, J. Wang, H. Liang, W. Zhang, H. Li, L. You, Z. Wang, L. Li, N.-L. Liu, K. Chen, T.-Y. Chen, and J.-W. Pan et al., Experimentally verified approach to nonentanglement-breaking channel certification, Phys. Rev. Lett. \textbf{124}, 010502 (2020).

\bibitem{zhen2020unified}Y.-Z. Zhen, Y. Mao, K. Chen, F. Buscemi, and O. Dahlsten, Unified approach to witness non-entanglement-breaking quantum channels, Phys. Rev. A \textbf{101}, 062301 (2020).

\bibitem{hsieh2017quantifying}J.-H. Hsieh, S.-H. Chen, and C.-M. Li, Quantifying quantum-mechanical processes, Sci. Rep. \textbf{7}, 13588 (2017).

\bibitem{kuo2019quantum}C.-C. Kuo, S.-H. Chen, W.-T. Lee, H.-M. Chen, H. Lu, and C.-M. Li, Quantum process capability, Sci. Rep. \textbf{9}, 20316 (2019).

\bibitem{rosset2018resource}D. Rosset, F. Buscemi, and Y.-C. Liang, Resource theory of quantum memories and their faithful verification with minimal assumptions, Phys. Rev. X \textbf{8}, 021033 (2018). % 96

\bibitem{chitambar2019quantum}E. Chitambar and G. Gour, Quantum resource theories, Rev. Mod. Phys. \textbf{91}, 025001 (2019). % 91

\bibitem{gour2019quantify}G. Gour and A. Winter, How to quantify a dynamical quantum resource, Phys. Rev. Lett. \textbf{123}, 150401 (2019). % 92

\bibitem{liu2019resource}Z.-W. Liu and A. Winter, Resource theories of quantum channels and the universal role of resource erasure, arXiv:1904.04201. % 95  % 8 April 2019

\bibitem{takagi2019general}R. Takagi and B. Regula, General resource theories in quantum mechanics and beyond: operational characterization via discrimination tasks, Phys. Rev. X \textbf{9}, 031053 (2019). % 98

\bibitem{theurer2019quantifying}T. Theurer, D. Egloff, L. Zhang, and M. B. Plenio, Quantifying operations with an application to coherence, Phys. Rev. Lett. \textbf{122}, 190405 (2019). % 100

\bibitem{hsieh2020resource}C.-Y. Hsieh, Resource preservability, Quantum \textbf{4}, 244 (2020). % 93

\bibitem{liu2020operational}Y. Liu and X. Yuan, Operational resource theory of quantum channels, Phys. Rev. Research \textbf{2}, 012035(R) (2020). % 94

\bibitem{saxena2020dynamical}G. Saxena, E. Chitambar, and G. Gour, Dynamical resource theory of quantum coherence, Phys. Rev. Research \textbf{2}, 023298 (2020). % 97

\bibitem{takagi2020application}R. Takagi, K. Wang, and M. Hayashi, Application of the resource theory of channels to communication scenarios, Phys. Rev. Lett. \textbf{124}, 120502 (2020). % 99

\bibitem{uola2020quantification}R. Uola, T. Kraft, and A. A. Abbott, Quantification of quantum dynamics with input-output games, Phys. Rev. A \textbf{101}, 052306 (2020). % 101

\bibitem{yuan2021universal}X. Yuan, Y. Liu, Q. Zhao, B. Regula, J. Thompson, and M. Gu, Universal and operational benchmarking of quantum memories, npj Quantum Inf. \textbf{7}, 108 (2021).  % 102

\bibitem{Liu20202020}Y. Liu, and X. Yuan, Operational resource theory of quantum channels, Phys. Rev. Research \textbf{2}, 012035(R) (2020).

\bibitem{Saxena20202020}G. Saxena, E. Chitambar, and G. Gour, Dynamical resource theory of quantum coherence, Phys. Rev. Research \textbf{2}, 023298 (2020).

\bibitem{chen2021quantifying}S.-H. Chen, M.-L. Ng, and C.-M. Li, Quantifying entanglement preservability of experimental processes, Phys. Rev. A \textbf{104}, 032403 (2021).

\bibitem{Zeilinger19981998}J.-W. Pan and A. Zeilinger, Greenberger-Horne-Zeilinger-state analyzer, Phys. Rev. A \textbf{57}, 2208 (1998).

\bibitem{Weinfurter20122012}J.-W. Pan, Z. B. Chen, C. Y. Lu, H. Weinfurter, A. Zeilinger, and M. Zukowski, Multiphoton entanglement and interferometry, Rev. Mod. Phys. \textbf{84}, 777 (2012).

\bibitem{ChuangPrescription}I. L. Chuang and M. A. Nielsen, Prescription for experimental determination of the dynamics of a quantum black box, J. Mod. Opt. \textbf{44}, 2455 (1997).

\bibitem{chia}C.-K. Chen, S.-H. Chen, N.-N. Huang, and C.-M. Li, Identifying genuine quantum teleportation, Phys. Rev. A \textbf{104}, 052429 (2021).

\bibitem{blais2007quantum}A. Blais, J. Gambetta, A. Wallraff, D. I. Schuster, S. M. Girvin, M. H. Devoret, and R. J. Schoelkopf, Quantum-information processing with circuit quantum electrodynamics, Phys. Rev. A \textbf{75}, 032329 (2007).

\bibitem{wendin2017quantum}G. Wendin, Quantum information processing with superconducting circuits: a review, Rep. Prog. Phys. \textbf{80}, 106001 (2017).

\bibitem{li2019tackling}G. Li, Y. Ding, and Y. Xie, Tackling the qubit mapping problem for NISQ-era quantum devices, in {\em Proceedings of the Twenty-Fourth International Conference on Architectural Support for Programming Languages and Operating Systems, 2019}, p. 1001.

\bibitem{bharti2021noisy}K. Bharti et al., Noisy intermediate-scale quantum algorithms, Rev. Mod. Phys. \textbf{94}, 015004 (2022).

\bibitem{steiner2000towards}M. Steiner, Towards quantifying non-local information transfer: finite-bit non-locality, Phys. Lett. A \textbf{270}, 239 (2000).

\bibitem{branciard2011quantifying}C. Branciard and N. Gisin, Quantifying the nonlocality of Greenberger-Horne-Zeilinger quantum correlations by a bounded communication simulation protocol, Phys. Rev. Lett. \textbf{107}, 020401 (2011).

\bibitem{kaszlikowski2000violations}D. Kaszlikowski, P. Gnaci{\'n}ski, M. {\.Z}ukowski, W. Miklaszewski, and A. Zeilinger, Violations of local realism by two entangled N -dimensional systems are stronger than for two qubits, Phys. Rev. Lett. \textbf{85}, 4418 (2000).

\bibitem{acin2002quantum}A. Ac{\'\i}n, T. Durt, N. Gisin, and J. I. Latorre, Quantum nonlocality in two three-level systems, Phys. Rev. A \textbf{65}, 052325 (2002). % 121

\bibitem{massar2002nonlocality}S. Massar, Nonlocality, closing the detection loophole, and communication complexity, Phys. Rev. A \textbf{65}, 032121 (2002). % 123

\bibitem{perez2008unbounded}D. P{\'e}rez-Garc{\'\i}a, M. M. Wolf, C. Palazuelos, I. Villanueva, and M. Junge, Unbounded violation of tripartite Bell inequalities, Commun. Math. Phys. \textbf{279}, 455 (2008). % 124

\bibitem{Discriminating2020} S.-H. Chen, H. Lu, Q.-C. Sun, Q. Zhang, Y.-A. Chen, and C.-M. Li, Discriminating quantum correlations with networking quantum teleportation, Phys. Rev. Research \textbf{2}, 013043 (2020).

\bibitem{mosek}MOSEK ApS, The mosek optimization toolbox for MATLAB version 9.3, \url{https://docs.mosek.com/9.3/toolbox/index.html}.
% MOSEK ApS, The MOSEK optimization toolbox for MATLAB manual. Version 8.0, 2017.  % https://journals.aps.org/prl/references/10.1103/PhysRevLett.128.204502

\bibitem{yalmip}J. L{\"o}fberg, Yalmip: A toolbox for modeling and optimization in MATLAB. In CACSD, 2004 IEEE International Symposium on Taipei, Taiwan. Available at \url{http://users.isy.liu.se/johanl/yalmip/}.

\bibitem{sdpt}K. C. Toh, M. J. Todd, and R. H. T{\"u}t{\"u}nc{\"u}, SDPT3 – a MATLAB software package for semidefinite-quadratic-linear programming, version 4.0. Available at \url{https://blog.nus.edu.sg/mattohkc/softwares/sdpt3/}.

\bibitem{HNN}N.-N. Huang, Master's thesis, National Cheng Kung University, 2020.

\bibitem{o2004quantum}J. L. O'Brien, G. J. Pryde, A. Gilchrist, D. F. V. James, N. K. Langford, T. C. Ralph, and A. G. White, Quantum process tomography of a controlled-NOT gate, Phys. Rev. Lett. \textbf{93}, 080502 (2004).

\bibitem{sourceCodeCH} \text{Qiskit, }\url{https://github.com/Qiskit/qiskit-aer/blob/main/qiskit_aer/noise/noise_model.py}.

\bibitem{awsCH} \text{Amazon Braket console, }\url{https://aws.amazon.com/tw/console/}.

\bibitem{rigettiCH} \text{Rigetti, }\url{https://www.rigetti.com/what-we-build}.

\bibitem{rigetti2CH} \text{Rigetti pyQuil, }\url{https://pyquil-docs.rigetti.com/en/latest/noise.html\#}.

\bibitem{Abrams2019}D. M. Abrams, N. Didier, B. R. Johnson, M. P. da Silva, and C. A. Ryan, Implementation of the XY interaction family with calibration of a single pulse, arXiv:1912.04424.

\bibitem{mangini2021qubit}S. Mangini, L. Maccone, and C. Macchiavello, Qubit noise deconvolution, EPJ Quantum Technol. \textbf{9}, 29 (2022).

\bibitem{rigetti3CH} \text{Quantum cloud services of Rigetti Computing, }\url{https://qcs.rigetti.com/lattices}.


\end{thebibliography}
\end{document}